\documentclass[11pt,a4paper]{article}
\usepackage[latin1]{inputenc} 
\usepackage{graphicx,latexsym,color,amsfonts,amsthm}
\usepackage[intlimits]{amsmath}
\usepackage{calc}
\usepackage{cite,citesort}

\usepackage{circuitikz}
\renewcommand{\vec}[1]{{\boldsymbol#1}}
\newif\ifgreek
\def\testgreek#1{
  \ifx#1\gamma\greektrue\else\ifx#1\Gamma\greektrue\else
  \ifx#1\epsilon\greektrue\else
  \ifx#1\mu\greektrue\else
  \ifx#1\rho\greektrue\else
  \ifx#1\sigma\greektrue\else\ifx#1\Sigma\greektrue\else
     \greekfalse
  \fi\fi\fi\fi\fi\fi\fi}

\newcommand{\mat}[1]{{\testgreek#1\ifgreek\boldsymbol#1\else
                      \mathbf#1\fi}} 

\newcommand{\ie}{\textit{i.e.}\/, }
\newcommand{\eg}{\textit{e.g.}\/, }
\newcommand{\cf}{\textit{cf.}\/, }

\providecommand*{\mrm}[1]{\mathrm{#1}}

\providecommand*{\eu}{\ensuremath{\mrm{e}}}

\renewcommand{\Re}{\operatorname{Re}}	
\renewcommand{\Im}{\operatorname{Im}}	
\providecommand*{\diff}{\operatorname{d}\!}

\providecommand*{\ju}{\ensuremath{\mrm{j}}}
\newcommand{\Ma}{_\mrm{m}}
\newcommand{\El}{_\mrm{e}}

\newcommand{\Jv}{\vec{J}}
\newcommand{\Ev}{\vec{E}}

\newcommand{\rv}{\vec{r}}
\newcommand{\We}{W\El}
\newcommand{\Wm}{W\Ma}

\newcommand{\psiv}{\vec{\psi}}
\newcommand{\qtext}[1]{\quad\text{#1}}
\newcommand{\herm}{\text{H}}
\newcommand{\Jm}{\mat{J}}

\newcommand{\Gm}{\mat{G}}
\newcommand{\Rm}{\mat{R}}

\newcommand{\norm}[1]{||#1||}

\newcommand{\Vm}{\mat{V}}
\newcommand{\Zm}{\mat{Z}}

\newcommand{\Hv}{\vec{H}}

\newcommand{\diffV}{\diff\mrm{V}}
							
\newcommand{\partder}[2]{\frac{\partial#1}{\partial{#2}}}

\renewcommand{\Jm}{\mat{I}}
\newcommand{\Jmt}{\tilde{\mat{I}}}

\newcommand{\Ym}{\mat{Y}}
\newcommand{\Xm}{\mat{X}}
\newcommand{\Bm}{\mat{B}}

\newcommand{\Um}{\mat{U}}
\newcommand{\Lambdam}{\mat{\Lambda}}

\newcommand{\QZmp}{Q_{\mathbf{Z}'}}
\newcommand{\QXmp}{Q_{\mathbf{X}'}}
\newcommand{\QZinp}{Q_{\mrm{Z'_{in}}}}

\newcommand{\QZp}{Q_{\mrm{Z'_{in}}}}

\newcommand{\QabsZp}{Q_{\mathbf{Z}'}}
\newcommand{\QabsYp}{Q_{\mathbf{Y}'}}

\newcommand{\QImZp}{Q_{\mathbf{X}'}}

\newcommand{\QXp}{Q_{\mathbf{X}'}}

\newcommand{\QZB}{Q_{\mrm{Z_{in}^{B}}}}
\newcommand{\WXpe}{W_{\mrm{e}\mathbf{X}'}}
\newcommand{\WXpm}{W_{\mrm{m}\mathbf{X}'}}
\newcommand{\WZpe}{W_{\mrm{e}\mathbf{Z}'}}
\newcommand{\WZpm}{W_{\mrm{m}\mathbf{Z}'}}
\newcommand{\refl}{\varGamma}

\newcommand{\tran}{\mrm{T}}

\newcommand{\Pv}{\vec{P}}
\newcommand{\Dv}{\vec{D}}

\newcommand{\clight}{\mrm{c}_0}
\newcommand{\epsiloninf}{\epsilon_\infty}
\newcommand{\epsilonr}{\epsilon_\mrm{r}}
\newcommand{\epsilons}{\epsilon_\mrm{s}}
\newcommand{\mur}{\mu_\mrm{r}}
\newcommand{\Zin}{Z_\mrm{in}}
\newcommand{\Rin}{R_\mrm{in}}
\newcommand{\Xin}{X_\mrm{in}}
\newcommand{\Yin}{Y_\mrm{in}}
\newcommand{\Gin}{G_\mrm{in}}
\newcommand{\Bin}{B_\mrm{in}}
\newcommand{\Vin}{V_\mrm{in}}
\newcommand{\Iin}{I_\mrm{in}}

\begin{document}

\title{Q factors for antennas in dispersive media}

\author{Mats Gustafsson\thanks{Department of Electrical and Information Technology, Lund University, Box 118, SE-221 00 Lund, Sweden. (Email: mats.gustafsson@eit.lth.se).}, Doruk Tayli, and Marius Cismasu}%

\maketitle

\begin{abstract}
Stored energy and Q-factors are used to quantify the performance of small antennas. Accurate and efficient evaluation of the stored energy is also essential for current optimization and the associated physical bounds. Here, it is shown that the frequency derivative of the input impedance and the stored energy can be determined from the frequency derivative of the electric field integral equation. The expressions for the differentiated input impedance and stored energies differ by the use of a transpose and Hermitian transpose in the quadratic forms. The quadratic forms also provide simple single frequency formulas for the corresponding Q-factors. The expressions are further generalized to antennas integrated in temporally dispersive media. Numerical examples that compare the different Q-factors are presented for dipole and loop antennas in conductive, Debye, Lorentz, and Drude media. The computed Q-factors are also verified with the Q-factor obtained from the stored energy in Brune synthesized circuit models. 
\end{abstract}

\section{Introduction}
Temporal dispersion is present in natural~\cite{Landau+Lifshitz1960,Jackson1999,VanBladel2007} and artificial materials~\cite{Engheta+Ziolkowski2006,Capolino2009a,Caloz2011}. Dispersion can often be neglected for antenna modeling in the microwave range but it is usually necessary for modeling of phenomena in the mm, THz, and optical range. Electromagnetic energy density in dispersive media builds on the classical results in~\cite{Landau+Lifshitz1960} with a renewed interest in applications such as; antennas, metamaterials, and photonics~\cite{Ruppin2002,Tretyakov2005,Vorobyev2012,Yaghjian+etal2013}. 

Antennas are often placed in the proximity of, or inside lossy media for example, submarines and body implants~\cite{Wheeler1958,Skrivervik2013}. The losses of the system are associated with conduction or relaxation of the media. These lead to a frequency dependent permittivity and hence temporal dispersion. Characterization of antennas in a lossy background medium is challenging as the electromagnetic fields decay exponentially away from the antenna and radiation patterns are coordinate dependent~\cite{Moore1963}. 

Stored energy is instrumental for antenna analysis in terms of the Q-factor. In~\cite{Yaghjian+Best2005,Yaghjian2007,Yaghjian+etal2013,Hansen+etal2014} stored energy is considered for small antennas composed of dispersive or lossy media, embedded in free space. The classical subtraction technique, where the energy in the far field is subtracted from the total energy density, is difficult to generalize to lossy media due to the exponential decay of the far field and its associated coordinate dependence. Using the power flow in the subtraction technique might be useful for spherical geometries,~\cite{Karlsson2004}. Here, we follow the approach by Harrington~\cite{Harrington1975}, Geyi~\cite{Geyi2003b}, and Vandenbosch~\cite{Vandenbosch2010} and express the stored energy and Q-factors in terms of the current density on the antenna structure. The derivation is based on frequency differentiation of the method of moments (MoM) impedance matrix. 

The Q-factor, $\QZp$, defined by the frequency derivative of the input impedance~\cite{Yaghjian+Best2005} is first expressed as a (bilinear) quadratic form in the current density. This simplifies the expression proposed in~\cite{Capek+etal2014} by eliminating the frequency derivatives of the current density and generalizes it to temporally dispersive media. Although, the $\QZp$ factor is inversely proportional to the fractional bandwidth~\cite{Yaghjian+Best2005,Gustafsson+Jonsson2014}, it is always possible to obtain $\QZp\approx 0$ with a simple matching network~\cite{Gustafsson+Nordebo2006b}. Therefore, care should be taken when using $\QZp$ for antenna optimization and deriving physical bounds. The quadratic form for the frequency derivative of the input impedance can also be useful for efficient interpolation of the input impedance over a frequency interval.

The stored energy expressions in~\cite{Vandenbosch2010} can produce negative values~\cite{Gustafsson+etal2012a} for large structures. This questions the validity of the energy expressions although several numerical tests indicate that the expressions are accurate for sub-wavelength antennas~\cite{Gustafsson+Jonsson2015a,Hazdra+etal2011}. The derivations in~\cite{Vandenbosch2010,Gustafsson+Jonsson2014,Vandenbosch2013a,Vandenbosch2013b} are based on subtraction of the radiated far field energy and hence not easily applicable for antennas in lossy media.
 
Here, we discuss some potential generalizations of the stored energy for antennas in general temporally dispersive media. These expressions are compared to the stored energy in circuit models that is determined from the input impedance using Brune synthesis~\cite{Wing2008,Gustafsson+Jonsson2015a}. The resulting Q-factors for dipole and loop antennas are compared for electric conduction, Debye, Lorentz, and Drude material models. The numerical results verify the expression for the differentiated Q-factor, $\QZp$. The results also indicate that the generalized expressions for the stored energy are valid for many cases of temporal dispersion. Strongly dispersive material models are used to investigate the validity of the expressions. Moreover, a particular material model with arbitrary small temporal dispersion is synthesized such that the Q-factor from the differentiated input impedance is negligible for self-resonant antennas. This indicates that it is difficult to express the stored energy solely in terms of the frequency derivative of the MoM impedance matrix for general temporally dispersive media.

This paper is organized as follows. In Sec.~\ref{S:DispMoM}, method of moments modeling of antennas in dispersive media is discussed. The Q-factor from the antenna input impedance $\QZp$ is derived in Sec.~\ref{S:Zinp}. Stored energy and the Q-factor for lumped circuit models are analyzed in Sec.~\ref{S:LumpedQ}. The stored energy and Q-factor for antennas are discussed in Sec.~\ref{S:antennaQ}. Numerical examples for dipoles and loops in conductive, Debye, Lorentz, and Drude models are shown in Sec.~\ref{S:NumEx}. 

\section{Antennas in temporally dispersive media}\label{S:DispMoM}
We consider antennas in a homogeneous temporally dispersive background medium. The antennas are modeled as perfect electric conductor (PEC) and the background medium has relative permittivity $\epsilonr(\omega)$ and relative permeability $\mur(\omega)$, where $\omega$ denotes the angular frequency. The wave impedance is $\eta=\sqrt{\mu/\epsilon}$ and the wavenumber is $k=-\ju\sqrt{-\omega^2\epsilon(\omega)\mu(\omega)}$.   

The impedance matrix is computed with the method of moments (MoM) formulation of the electric field integral equation (EFIE) using the Galerkin procedure~\cite{Peterson+Ray+Mittra1998}.
 The basis functions are assumed to be real valued, divergence conforming, with vanishing normal components at the antenna boundary~\cite{Peterson+Ray+Mittra1998}. A standard MoM implementation of the EFIE determines the impedance matrix $\Zm=\Rm+\ju\Xm$, with the elements
\begin{equation}
	\frac{Z_{mn}}{\eta} 
	=\ju\int_{V}\!\!\int_{V}\left(
  k^2\psiv_{m1}\cdot\psiv_{n2}\right.
  \left.-\nabla_1\cdot\psiv_{m1}\nabla_2\cdot\psiv_{n2}\right)
\frac{\eu^{-\ju k R_{12}}}{4\pi kR_{12}} 
	\diffV_1\diffV_2,
\label{eq:EFIE}
\end{equation} 
where $\psiv_{ni}$ is a short hand notation for basis functions $\psiv_n(\rv_i)$ with $n=1,...,N$ and $i=1,2$, $\rv$ denotes the position vector, and $R_{12}=|\rv_1-\rv_2|$. The current column matrix $\Jm$ contains the expansion coefficients $I_n$ for the
current density $\Jv(\rv)=\sum_{n=1}^N I_n\psiv_n(\rv)$ that is determined from the linear system
\begin{equation}
  \Zm\Jm=\Vm
  \qtext{or }
  \Jm = \Zm^{-1}\Vm=\Ym\Vm,
\label{eq:Jm}
\end{equation} 
where $\Vm$ is the column matrix with excitation coefficients and $\Ym=\Gm+\ju\Bm$ is the admittance matrix. 
The MoM impedance matrix in temporally dispersive media~\eqref{eq:EFIE} is formally identical to the free space case with the use of the complex-valued wavenumber $k$ in the background medium and normalization to the complex-valued background impedance $\eta$. 

The input voltage $V_{\mrm{in}}$ can be chosen freely and is here assumed real valued and frequency independent. This a real valued $\Vm$ with non-zero elements corresponding to the input voltage $V_{\mrm{in}}$. The input impedance, $\Zin=\Rin+\ju\Xin=\Yin^{-1}$, is determined from the admittance matrix
\begin{equation}
   \Zin=\frac{1}{\Yin}
  = \frac{V_{\mrm{in}}^2}{\Vm^{\tran}\Ym\Vm},
 \label{eq:Zin}
\end{equation}
where $\Yin=\Gin+\ju\Bin$ is the input admittance, for a single port antenna. The Q-factor for an antenna tuned to resonance is defined as
\begin{equation}
	Q = \frac{2\omega\max\{\We,\Wm\}}{P_{\mrm{d}}},
\label{eq:Qfactor}
\end{equation}
where $\We$ and $\Wm$ denote the stored electric and magnetic energies and $P_{\mrm{d}}$ is the dissipated power.
The dissipated power is determined from the Poynting vector and can be written~\cite{Pozar1983,Geyi2011,Vandenbosch2010,Gustafsson+Jonsson2014}
\begin{equation}
P_{\mrm{d}}
=\frac{1}{2}\Re\{\Jm^{\herm}\Vm\}
=\frac{1}{2}\Jm^{\herm}\Rm\Jm
=\frac{1}{2}\Vm^{\herm}\Gm\Vm.
\label{eq:}
\end{equation}

\section{Frequency derivative of the input impedance}\label{S:Zinp}
For a self-resonant single resonance antenna, we have the $\QZp$ estimate for the fractional bandwidth~\cite{Yaghjian+Best2005}
\begin{equation}\label{eq:Q2B}
	B \approx \frac{2}{\QZp}\frac{\refl_0}{\sqrt{1-\refl_0^2}},
\end{equation}
where $\refl_0$ denotes the threshold of the reflection coefficient, $\refl$, and
\begin{equation}
  \QZp = \frac{\omega|\Zin'|}{2\Rin}
  =\frac{\omega|\Yin'|}{2\Gin}
  =\omega|\refl'|
\label{eq:QZinp}
\end{equation}
where $\Zin'$, $\Yin'$, and $\refl'$ are the input impedance, input admittance, and  reflection coefficient derivatives, respectively~\cite{Yaghjian+Best2005}. The $\QZp$ was first expressed in terms of the current density $\Jv$ and its frequency derivative $\Jv'$ in~\cite{Capek+etal2014}. Here, we follow~\cite{TEAT-7231} and use the EFIE impedance matrix~\eqref{eq:EFIE} to express $\QZp$ solely in the current $\Jm$ or equivalently $\Jv$. 

We use the frequency derivative of the impedance matrix~\eqref{eq:EFIE}
to express $\QZp$~\eqref{eq:QZinp} in the current. The (angular) frequency derivative of the admittance matrix is
\begin{equation}
  \Ym'=\partder{\Ym}{\omega}=\partder{\Zm^{-1}}{\omega}=-\Zm^{-1}\Zm'\Zm^{-1}
  =-\Ym\Zm'\Ym.
\label{eq:}
\end{equation}
Using a real-valued frequency independent input voltage $V_{\mrm{in}}$ implying a voltage source $\Vm'=\mat{0}$ in~\eqref{eq:Jm}, we get the frequency derivative of the input admittance $Y_{\mrm{in}}$ from
\begin{equation}
  V_{\mrm{in}}^2Y_{\mrm{in}}'
  = (\Vm^{\tran}\Ym\Vm)'
  =\Vm^{\tran}\Ym'\Vm
  =-\Jm^{\tran}\Zm'\Jm.
\label{eq:izi}
\end{equation}
Here we note that~\eqref{eq:izi} is valid for frequency dependent currents, $\Jm$, and that the derivation is solely based on a frequency independent input voltage $V_{\mrm{in}}$. Moreover, the current on the antenna structure, $\Jm$, is in-general frequency dependent, \ie $\Jm'=\Ym'\Vm\neq\mat{0}$ as the admittance matrix, $\Ym$, is frequency dependent. This differs from the expression in~\cite{Capek+etal2014} that contains frequency derivatives of the current and is based on assuming a frequency independent input current, see also~\cite{Yaghjian+Best2005,Geyi2014}. 
The assumption of a real-valued frequency independent voltage source also agrees with modeling of a voltage gap in the MoM~\cite{Peterson+Ray+Mittra1998}. The corresponding frequency derivative of the input impedance is $Z_{\mrm{in}}'=-Z_{\mrm{in}}^2Y_{\mrm{in}}'$. 
The Q-factor~\eqref{eq:QZinp} defined from the frequency derivative of the input admittance and input impedance are evaluated using
\begin{equation}
  \frac{\omega \Yin'}{2\Gin}
  =\frac{\omega \Vin^2\Yin'}{2 \Vin^2 \Gin}
  =-\frac{\omega\Jm^{\tran}\Zm'\Jm }{2\Jm^{\herm}\Rm\Jm}
\label{eq:QYinpJ}
\end{equation}
and
\begin{equation}
  \frac{\omega Z_{\mrm{in}}'}{2R_{\mrm{in}}}
  =\frac{-\omega Z_{\mrm{in}}^2 Y_{\mrm{in}}'}{2R_{\mrm{in}}}
  =\frac{\omega |Y_{\mrm{in}}^2|\Jm^{\tran}\Zm'\Jm}{2Y_{\mrm{in}}^2\Jm^{\herm}\Rm\Jm},
\label{eq:QZinpJ}
\end{equation}
respectively, giving
\begin{equation}
  \QZp = \frac{\omega|\Jm^{\tran}\Zm'\Jm|}{2\Jm^{\herm}\Rm\Jm}
\label{eq:QZinpJ1}
\end{equation}
for self-resonant antennas.

We are most interested in the Q-factor of antennas tuned to resonance. The frequency derivative depends on the used matching network~\cite{Gustafsson+Nordebo2006b}. For series tuning with a lumped capacitor or inductor we have the Q-factor~\cite{Yaghjian+Best2005}
\begin{multline}
    \QZp^{\mrm{(s)}}
  =\left|\frac{\omega \Zin'}{2\Rin}+\ju\frac{|\Xin|}{2\Rin}\right|
  =\frac{\sqrt{(\omega \Rin')^2+(\omega \Xin'+|\Xin|)^2}}{2R_{\mrm{in}}}\\
  =\left|\frac{\omega \Yin'|\Yin^2|}{2\Gin \Yin^2}-\ju\frac{|\Bin|}{2\Gin}\right|
\label{eq:QZps} 
\end{multline}
and the case with parallel tuning elements is
\begin{equation}
  \QZp^{\mrm{(p)}} =\left|\frac{\omega \Yin'}{2\Gin}+\ju\frac{|\Bin|}{2\Gin}\right|
  =\left|\frac{\omega \Zin'|\Zin^2|}{2\Rin \Zin^2}-\ju\frac{|\Xin|}{2\Rin}\right|.
\label{eq:QZpp}
\end{equation}
The series case~\eqref{eq:QZps} is most commonly used~\cite{Yaghjian+Best2005}. However, the two tuning cases $\QZp^{\mrm{(s)}}$ and $\QZp^{\mrm{(p)}}$ are similar and here we consider the maximal value of $\QZp^{\mrm{(s)}}$ and $\QZp^{\mrm{(p)}}$ to define
\begin{equation}
  \QZp = \max\{\QZp^{\mrm{(s)}},\QZp^{\mrm{(p)}}\}.
\label{eq:QZYinp}
\end{equation}
This definition removes the ambiguity of the tuning element or equivalently the preference of the input impedance or input admittance. The practical difference is often small but sometimes observable around the resonance and anti-resonance frequencies. The tuning factor in~\eqref{eq:QZps} and \eqref{eq:QZpp} is the difference between the stored magnetic and electric energies normalized with the dissipated power and can be written in different forms,~\eg
\begin{equation}
  \frac{\omega|\Wm-\We|}{2P_{\mrm{d}}}
  =\frac{|\Xin|}{2\Rin} = \frac{|\Bin|}{2\Gin} 
  =\frac{|\Jm^{\herm}\Xm\Jm|}{2\Jm^{\herm}\Rm\Jm} 
  =\frac{|\Vm^{\herm}\Bm\Vm|}{2\Vm^{\herm}\Gm\Vm}. 
\label{eq:tuningfactor}
\end{equation}

The frequency derivative of the EFIE impedance matrix $\Zm$ in~\eqref{eq:EFIE} is
\begin{multline}
  \omega\partder{Z_{mn}}{\omega}
  =\omega\eta\partder{(Z_{mn}/\eta)}{\omega}
  +\omega\frac{Z_{mn}}{\eta}\partder{\eta}{\omega}\\
  =k\partder{(Z_{mn}/\eta)}{k}\frac{\eta\omega}{k}\partder{k}{\omega}
  +\omega\frac{Z_{mn}}{\eta}\partder{\eta}{\omega}
\label{eq:EFIEp1}
\end{multline}
for a temporally dispersive background medium with $k=\omega\sqrt{\epsilon\mu}$, $\eta=\sqrt{\mu/\epsilon}$, $k\eta=\omega\mu$, and $k/\eta=\omega\epsilon$. For the common case of a non-magnetic medium, $\mur=1$, the result simplifies to 
\begin{equation}
  \omega\partder{Z_{mn}}{\omega} 
=k\partder{(Z_{mn}/\eta)}{k}\eta\left(\frac{\omega\partial\epsilon}{2\epsilon\partial\omega}+1\right)
  -\frac{Z_{mn}}{2}\frac{\omega\partial\epsilon}{\epsilon\partial\omega}.
\label{eq:EFIEp}
\end{equation}
The terms~\eqref{eq:EFIEp1} and~\eqref{eq:EFIEp} involve the impedance matrix $\Zm$, its wavenumber derivative, and frequency derivatives of the material parameters.  
The differentiation with respect to the background wavenumber $k$ of the EFIE impedance matrix~\eqref{eq:EFIE} normalized with the background impedance is
\begin{multline}
  k\partder{}{k}\frac{Z_{mn}}{\eta} 
  	=\int_{V}\!\!\int_{V} 
    \Big(\ju
  (k^2\psiv_{m1}\cdot\psiv_{n2}
  +\nabla_1\cdot\psiv_{m1}\nabla_2\cdot\psiv_{n2})
  \\
  +(k^2\psiv_{m1}\cdot\psiv_{n2}
  -\nabla_1\cdot\psiv_{m1}\nabla_2\cdot\psiv_{n2})kR_{12}
	\Big)
  \frac{\eu^{-\ju k R_{12}}}{4\pi kR_{12}}\diffV_1\diffV_2,
\label{eq:EFIEkp}
\end{multline}
where we observe that the first term resembles the impedance matrix~\eqref{eq:EFIE} but with an addition instead of subtraction of the two terms. 
The second term is non-singular due to the multiplication with $kR_{12}$.
Combining~\eqref{eq:EFIEp} and~\eqref{eq:EFIEp1} with~\eqref{eq:QZps}, \eqref{eq:QZpp} and~\eqref{eq:QYinpJ} expresses $\QZp$ as a quadratic form in the current  $\Jm$. The corresponding expressions in~\cite{Capek+etal2014} differ from~\eqref{eq:QZinpJ1} as they include frequency derivatives and complex conjugates of the current density.

The frequency derivative~\eqref{eq:EFIEkp} involves both the real and imaginary part of the impedance matrix. A series expansion in the wavenumber reveals that
\begin{equation}\label{eq:Zmp_kseries}
  \frac{k\,\partial R_{mn}}{\eta_0\,\partial k} 
  \sim k^2
  \qtext{and }
  \frac{k\,\partial X_{mn}}{\eta_0\,\partial k} 
  \sim k^{-1}  
\end{equation}
as $k\to 0$ and hence the derivative of the reactance dominates for small antennas.

The matrix $\Zm'$ is symmetric and can hence be Takagi factorized as $\Zm'=\Um^{\tran}\Lambdam\Um$, where $\Lambdam$ is a diagonal matrix containing the eigenvalues (non-negative) of $\Zm'\Zm^{\prime\herm}$. This gives the quadratic (bilinear) form 
\begin{equation}
  \Jm^{\tran}\Zm'\Jm
  =(\Um\Jm)^{\tran}\Lambdam\Um\Jm
  =\Jmt^{\tran}\Lambdam\Jmt
  =\sum_{n=1}^N \tilde{I}_n^2\lambda_n,
\label{eq:Zmp_Takagi}
\end{equation}
where we note that it is always possible to find non-zero currents $\Jm$ such that $\Jm^{\tran}\Zm'\Jm=0$ if there are at least two non-zero eigenvalues (or modes) $\lambda_n$. This shows that the lower bound on $\QZp$ is in general 0 if the currents are chosen arbitrary, see also the explicit construction using matching circuits in~\cite{Gustafsson+Nordebo2006b}. This problem with $\QZp$ stems from the use of the transpose of $\Jm$ in the (bilinear) quadratic form~\eqref{eq:QZinpJ1}. The corresponding energy expressions are formulated as quadratic (sesquilinear) forms involving the Hermitian transpose of $\Jm$ giving positive semidefinite forms suitable for optimization~\cite{Gustafsson+Nordebo2013,Cismasu+Gustafsson2014a,Cismasu+Gustafsson2014a}.

\section{Q in lumped circuit models of antennas}\label{S:LumpedQ} 
The input impedance can be used to synthesize a lumped circuit model of the antenna that is used to determine the antenna Q~\cite{Chu1948,Thal2006,Thal2012,Gustafsson+Jonsson2015a}. Consider a lumped circuit network with resistors, inductors, and capacitors. The input impedance between two nodes of the network is defined by the quotient between the voltage and current. The circuit is fed using either a voltage or current source between the nodes. The results are related and correspond to an interchange between the impedance and admittance in the discussion below. For simplicity, we assume a voltage source and use the Kirchhoff voltage law to construct the linear system~\cite{Guillemin1963}
\begin{equation}
  \Zm\mat{I} = \Vm,
\label{eq:lumped_ZIV}
\end{equation}
where the impedance matrix $\Zm=\Rm+\ju\Xm$ contains elements of the form $Z_{mn}=R_{mn}+\ju\omega L_{mn}-\ju/(\omega C_{mn})$ depending on the lumped elements in the branch. Order the branches such that the voltage matrix $\Vm$ contains the source voltage $\Vin$ at one position and is zero elsewhere. The corresponding current matrix contains the input current at the same position. The input impedance $\Zin=\Vin/\Iin$ and input admittance $\Yin=1/\Zin$ are determined from~\cite{Guillemin1963},
\begin{equation}
  \Yin\Vin^2=\Zin\Iin^2=\Vin\Iin=
  \Vm^{\tran}\mat{I}=\mat{I}^{\tran}\Zm\mat{I}  
\label{eq:lumped_Zin}
\end{equation}
that can also be derived using Tellegan's theorem~\cite{Wing2008}.
The voltage source is frequency independent, $\Vin'=0$, so differentiation of the input admittance~\eqref{eq:lumped_Zin} with respect to the frequency gives 
\begin{equation}
  \Yin'\Vin^{2}
  =\Vm^{\tran}\mat{I}'
  =\mat{I}^{\tran}\Zm\mat{I}'
  =-\mat{I}^{\tran}\Zm'\mat{I} 
  =-\ju\mat{I}^{\tran}\Xm'\mat{I},   
\label{eq:lumped_Zinp}
\end{equation}
where differentiation of~\eqref{eq:lumped_ZIV} is used to get $\Zm'\mat{I}+\Zm\mat{I}'=\mat{0}$. Moreover, the resistance matrix $\Rm$ is frequency independent so the differentiated impedance matrix $\Zm'=\ju\Xm'$ is imaginary valued with the elements 
\begin{equation}
  X_{mn}' = 
  \partder{}{\omega}\left(\omega L_{mn}-\frac{1}{\omega C_{mn}}\right) = L_{mn} + \frac{1}{\omega^2 C_{mn}}.
\label{eq:lumped_Xp}
\end{equation}
Here, it is important to realize that the current matrix $\mat{I}$ in~\eqref{eq:lumped_Zinp} is complex valued and hence the differentiated input admittance $\Yin'$ is complex valued. The differentiated input impedance is finally $\Zin'=-\Zin^2\Yin'$. Moreover, we observe that the expression is similar to the corresponding expression for the antenna input impedance~\eqref{eq:izi}. The frequency differentiated impedance matrix for the antenna case can have a non-zero contribution from the resistance matrix. The Q-factors from the differentiated input impedance are analogous to the antenna case~\eqref{eq:QZinpJ1}.

The Q-factor definition in~\eqref{eq:Qfactor} includes the stored electric $\We$ and magnetic $\Wm$ energies. The stored energies~\cite{Wing2008,Guillemin1963} in capacitors and inductors are $|V|^2C/4=|I|^2/(4\omega^2C)$ and $|I|^2L/4$, respectively. Comparing the stored energy in capacitors and inductors with~\eqref{eq:lumped_Xp} shows that the frequency derivative of the impedance matrix gives the total stored energy
\begin{equation}
  \We+\Wm = \frac{\mat{I}^{\herm}\Xm'\mat{I}}{4}\geq 0
\label{eq:lumped_Wem}
\end{equation}
that resembles~\eqref{eq:lumped_Zinp} with the transpose replaced by a Hermitian transpose.
The difference between the stored magnetic and electric energies is $\Wm-\We = \frac{1}{4\omega}\mat{I}^{\herm}\Xm\mat{I}$, giving the stored magnetic and electric energies as
\begin{equation}
  \Wm  
  =\frac{1}{8}\mat{I}^{\herm}\left(\partder{\Xm}{\omega}+\frac{\Xm}{\omega}\right)\mat{I}
  =\frac{1}{4}\sum_{m,n=1}^N I_m^{\ast}L_{mn}I_n\geq 0
\label{eq:WmI}
\end{equation}
and
\begin{equation}
  \We  
  =\frac{1}{8}\mat{I}^{\herm}\left(\partder{\Xm}{\omega}-\frac{\Xm}{\omega}\right)\mat{I} 
    =\frac{1}{4\omega^2}\sum_{m,n=1}^N I_m^{\ast}C^{-1}_{mn}I_n\geq 0,
\label{eq:WeI}
\end{equation}
respectively. The dual formulation of~\eqref{eq:WmI} and \eqref{eq:WeI}, using a current source changes currents to voltages and the impedance matrix to the admittance matrix, \eg $\We+\Wm=\Vm^{\herm}\Bm'\Vm/4$.

The frequency differentiated reactance matrix $\Xm'$ is real valued symmetric positive semi definite and can be diagonalized as $\Xm'=\Um^{\tran}\Lambdam\Um$, where $\Lambdam$ is a diagonal matrix containing the eigenvalues (non-negative) and $\Um$ is a real-valued unitary matrix. This gives an inequality between the expression for the frequency derivative and the stored energy
\begin{equation}
  \mat{I}^{\herm}\Xm'\mat{I}
  =(\Um\mat{I})^{\herm}\Lambdam\Um\mat{I}
  \geq 
  \left|(\Um\mat{I})^{\tran}\Lambdam\Um\mat{I}\right|
  =\left|\mat{I}^{\tran}\Xm'\mat{I}\right|.
\label{eq:lumped_Xp_inequality}
\end{equation}
The inequality becomes an equality for currents with a constant phase, \eg real-valued currents. This condition for equality is sufficient but not necessary as seen by interchanging the impedance and admittance formulations. For a self-resonant input impedance, $\Xin=0$, this shows that the $Q$ determined from the stored energy is always greater than or equal to the $Q$ determined from the frequency derivative, \ie
\begin{equation}
  Q = \frac{\omega(\We+\Wm)}{P_{\mrm{d}}}
  =\frac{\omega\mat{I}^{\herm}\Xm'\mat{I}}{2\mat{I}^{\herm}\Rm\mat{I}}
  \geq \frac{\omega|\mat{I}^{\tran}\Xm'\mat{I}|}{2\mat{I}^{\herm}\Rm\mat{I}}
  =\QZp.
\label{eq:}
\end{equation}

For an input impedance tuned to resonance, we have
\begin{multline}
  Q = \frac{2\omega\max\{\We,\Wm\}}{P_{\mrm{d}}}
  =\frac{\max\{\mat{I}^{\herm}(\omega\Xm'\pm\Xm)\mat{I}\}}{2\mat{I}^{\herm}\Rm\mat{I}}
  =\frac{\omega\mat{I}^{\herm}\Xm'\mat{I}+|\mat{I}^{\herm}\Xm\mat{I}|}{2\mat{I}^{\herm}\Rm\mat{I}}\\
  \geq \frac{|\omega\mat{I}^{\tran}\Xm'\mat{I}|+|\mat{I}^{\herm}\Xm\mat{I}|}{2\mat{I}^{\herm}\Rm\mat{I}}
  = \max_{|\alpha|=1}\frac{|\omega\mat{I}^{\tran}\Xm'\mat{I}\alpha|+|\mat{I}^{\herm}\Xm\mat{I}|}{2\mat{I}^{\herm}\Rm\mat{I}}
  \geq \max_{|\alpha|=1}\frac{|\omega\mat{I}^{\tran}\Xm'\mat{I}\alpha+\mat{I}^{\herm}\Xm\mat{I}|}{2\mat{I}^{\herm}\Rm\mat{I}} \\
  \geq\max\left\{\frac{|\omega\Yin'\pm\ju|\Bin||}{2\Gin},\frac{|\omega\Zin'\pm\ju|\Xin||}{2\Rin}\right\}
  \geq \QZp,
\label{eq:lumped_Qtuned}
\end{multline}
where we used the series~\eqref{eq:QZps} and parallel~\eqref{eq:QZpp} tuning elements with $\alpha=\{\pm 1, \pm |\Yin|^2/\Yin^2\}$. This is a general inequality between the Q-factors derived from the differentiated impedance and the stored energy valid for lumped circuit networks. Here, we investigate its implications for the stored energy expressed as the current density on the antenna.

\section{Q and stored energy for antennas}\label{S:antennaQ}
The frequency derivative of the impedance matrix~\eqref{eq:EFIEp} together with the lumped circuit expressions~\eqref{eq:lumped_Zinp} and~\eqref{eq:lumped_Wem} also shed new light on the stored energy expressions derived by Vandenbosch~\cite{Vandenbosch2010} and proposed by Harrington~\cite{Harrington1975}. 
The frequency derivative of the reactance matrix $\Xm'$ produces the quadratic forms for the stored energies in~\cite{Vandenbosch2010,Gustafsson+Jonsson2014,Cismasu+Gustafsson2014a}, \ie 
\begin{equation}
  \WXpe+\WXpm = \frac{1}{4}\Jm^{\herm}\Xm'\Jm. 
\label{eq:WemJV}
\end{equation}
This is a Hermitian quadratic form in terms of the frequency derivative of the reactance matrix. The difference between the stored magnetic and electric energies gives the explicit formulas for the stored magnetic and electric energies
\begin{equation}
  \WXpm  
  =\frac{1}{8}\Jm^{\herm}\left(\partder{\Xm}{\omega}+\frac{\Xm}{\omega}\right)\Jm 
\label{eq:WmJV}
\end{equation}
and
\begin{equation}
  \WXpe  
  =\frac{1}{8}\Jm^{\herm}\left(\partder{\Xm}{\omega}-\frac{\Xm}{\omega}\right)\Jm,   
\label{eq:WeJV}
\end{equation}
respectively.
The relations~\eqref{eq:WmJV} and~\eqref{eq:WeJV} are formally identical to the stored energy expressions for the lumped circuit networks~\eqref{eq:WmI} and~\eqref{eq:WeI}. They also resemble the expressions in~\cite{Harrington1968,Geyi+Jarmuszewski+Qi2000} for the input impedance of single and array antennas. Here, it is essential to note that~\eqref{eq:WmJV} and~\eqref{eq:WeJV} are expressed in the EFIE impedance matrix and not in the input impedance, \cf~\cite{Harrington1968,Geyi+Jarmuszewski+Qi2000}. Geyi~\cite{Geyi2014} has also recently proposed modifications involving frequency derivatives of the current density.

The Q-factor for antennas tuned to resonance~\eqref{eq:Qfactor} is 
\begin{equation}
  \QXp = 
   \frac{\max\{\Jm^{\herm}(\omega\Xm'\pm\Xm)\Jm\}}{2\Jm^{\herm}\Rm\Jm}
  =\frac{\omega\Jm^{\herm}\Xm'\Jm+|\Jm^{\herm}\Xm\Jm|}{2\Jm^{\herm}\Rm\Jm}
\label{eq:QXp}
\end{equation}
for the stored energy~\eqref{eq:WemJV}. Although the expressions for the frequency derivative~\eqref{eq:QZinpJ} and stored energy~\eqref{eq:WemJV} are similar, there are some fundamental differences. For a real valued (with a constant phase) current the stored energies only differ by the frequency derivative of the resistance $\Rm'$.
This implies that $\QZp\geq \QXp$ for real-valued currents due to this non-zero $\Rm'\neq \mat{0}$. This differs from the typical numerical results~\cite{Gustafsson+Jonsson2015a} and is explained by the in many cases negligible $\Rm'$, \ie $\norm{\Rm'}\ll \norm{\Xm'}$, see~\eqref{eq:Zmp_kseries}. For small self-resonant antennas with non-constant currents and a negligible $\Rm'$, \ie $\norm{\Rm'}\approx 0$, we have $\QZp\leq \QXp$ as $|\Jm^{\tran}\Xm'\Jm|\leq \Jm^{\herm}\Xm'\Jm$ for symmetric real-valued positive semi-definite matrices $\Xm'$ in accordance with the lumped circuit case~\eqref{eq:lumped_Xp_inequality}.

One problem with expression~\eqref{eq:WemJV} for the stored energy is that it can produce negative values for large structures~\cite{Gustafsson+etal2012a}. This questions the validity of the expression~\eqref{eq:WemJV}. Moreover, the inequality between the frequency derivative and the stored energy for lumped circuits~\eqref{eq:lumped_Xp_inequality} is in general not satisfied with~\eqref{eq:WemJV}.
An alternative expression for the stored energy that is non-negative and resembles the expressions from the differentiated input impedance~ \eqref{eq:QZinpJ} is given by
\begin{equation}
  \WZpe+\WZpm \approx \frac{1}{4}\left|\Jm^{\herm}\Zm'\Jm\right|.
\label{eq:}
\end{equation}
For small antennas the frequency derivative of the resistance is negligible compared with the frequency derivative of the reactance~\eqref{eq:Zmp_kseries} and the frequency derivative of the reactance is positive semi-definite. This simplifies the energy to the expressions~\eqref{eq:WemJV} introduced by Vandenbosch~\cite{Vandenbosch2010}, \ie
\begin{equation}
  \WZpe+\WZpm \approx \frac{1}{4}\left|\Jm^{\herm}\Zm'\Jm\right|
  \approx \frac{1}{4}\Jm^{\herm}\Xm'\Jm
	=\WXpe+\WXpm
\label{eq:}
\end{equation}
for $ka\ll 1$, where $a$ denotes the radius of the smallest sphere that circumscribes the antenna.
For antennas tuned to resonance we can use the energy difference~\eqref{eq:tuningfactor} to get the Q-factor
\begin{equation}
  \QabsZp = \frac{\omega|\Jm^{\herm}\Zm'\Jm| + |\Jm^{\herm}\Xm\Jm|}{2\Jm^{\herm}\Rm\Jm},
\label{eq:QZp}
\end{equation}
similar to the tuned case for lumped circuits~\eqref{eq:lumped_Qtuned}. The expression is also equal to the result from the frequency derivative of the input impedance for real valued currents $\Jm$. 

We can also use the admittance to introduce the Q-factor in an alternative form,
\begin{equation}
  \QabsYp = \frac{\omega|\Vm^{\herm}\Ym'\Vm| + |\Vm^{\herm}\Bm\Vm|}{2\Vm^{\herm}\Gm\Vm}
\label{eq:QYp}
\end{equation}
Here, we note that for a real valued frequency independent voltage source $\Vm^{\herm}\Ym'\Vm=\Vm^{\tran}\Ym'\Vm=-\Jm^{\tran}\Zm'\Jm$, giving $\QabsYp\geq \QZp$ for the self-resonant case.

For current optimization~\cite{Gustafsson+Nordebo2013,Cismasu+Gustafsson2014a,Cismasu+Gustafsson2014b,Gustafsson+etal2014c} it is essential to have expressions of the stored energy that are convex in the current density. This directly eliminates the expressions for the differentiated input impedance as they are non-convex due to the transpose giving a bilinear quadratic form. Also expressions involving the frequency derivative of the impedance or admittance matrix are in general non-convex \eg due to the indefinite sign of $\Rm'$.

\section{Numerical examples}\label{S:NumEx}
We compare the proposed Q-factors $\QZinp$ in~\eqref{eq:QZYinp}, $\QZB$ in~\eqref{eq:lumped_Qtuned}, $\QXmp$ in~\eqref{eq:QXp}, and $\QZmp$ in~\eqref{eq:QZp}. The circuit model for $Q=\QZB$ in~\eqref{eq:lumped_Qtuned} is determined using Brune synthesis from the input impedance of a single port antenna~\cite{Wing2008,Brune1931,Gustafsson+Jonsson2015a}. The antenna parameters are computed using a MoM code based on rectangular elements and divergence conforming basis functions~\cite{Peterson+Ray+Mittra1998} for planar (negligible thickness) structures modeled as a perfect electrical conductor (PEC). 

We consider the conductivity, Debye, Lorentz, and Drude material models for the permittivity~\cite{Jackson1999,Landau+Lifshitz1960,VanBladel2007} together with a free space permeability. 
To conclude, an example dispersion model with parameters arbitrary close to free space, giving negligible $\Zm'$, is synthesized. This example illustrates the difficulties to derive universally valid expressions for Q-factors. The Q-factors are depicted for dipole and loop antennas. The dipole antenna has length $\ell$, width $0.01\ell$, and is either fed in the center or $0.27\ell$ from the center. The loop antenna has height $\ell$, width $0.5\ell$, and a strip width of $\ell/64$ and is fed in the center of the longer side.

The results are presented in the dimensionless parameter $\ell/\lambda$, where $\lambda$ is the free-space wavelength. The material parameters are functions of the dimensionless parameter $\omega=2\pi\ell/\lambda$.

\subsection{Conductivity model}
We consider a strip dipole fed $0.27\ell$ from the center in a homogenous medium with relative permittivity $\epsilonr=1-\ju\sigma/\omega$, where $\omega=2\pi\ell/\lambda$. The off-center feed is chosen to eliminate some of the symmetries of the induced current density distribution in comparison to the center fed case, and increase the phase shift of the induced current density. 

\begin{figure}[t]
\begin{center}
\noindent
  \includegraphics[width=0.68\linewidth]{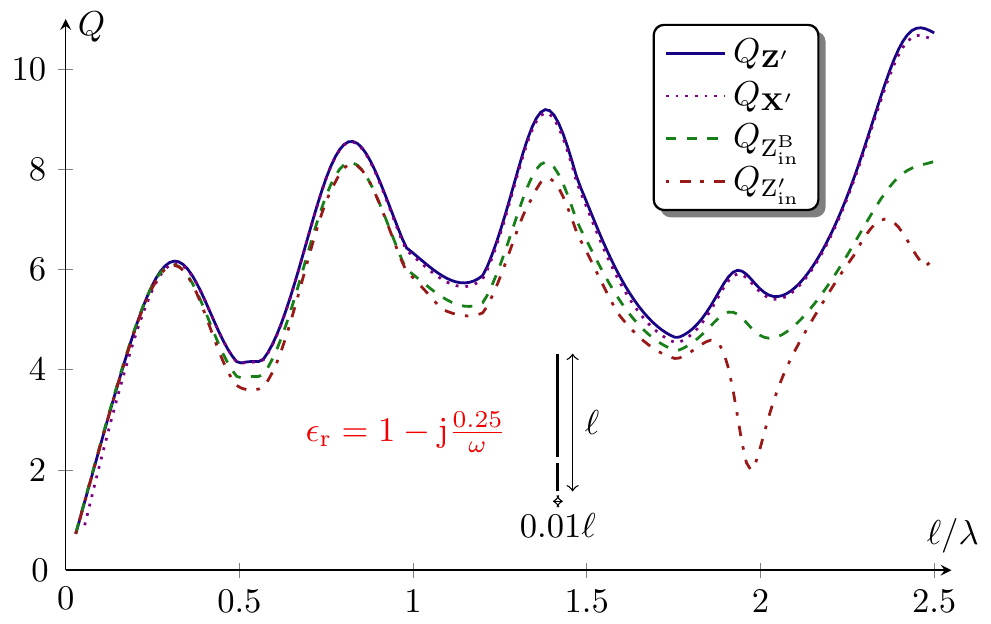}
  \caption{Q-factors for a strip dipole with length $\ell$, width $\ell/100$, fed $0.27\ell$ from the center, and placed in a homogeneous medium with relative permittivity $\epsilonr=1-0.25\ju/\omega$, where $\omega=2\pi\ell/\lambda$. The Q-factors are determined from the Brune synthesized circuit model $\QZB$ as in~\cite{Gustafsson+Jonsson2015a}, the differentiated input impedance $\QZinp$ in~\eqref{eq:QZYinp}, frequency derivative of the reactance matrix $\QXmp$ in~\eqref{eq:QXp}, and the frequency derivative of the impedance matrix $\QZmp$ in~\eqref{eq:QZp}.}
  \label{fig:dipole_c0p27_cond0_Q}
\end{center}
\end{figure}

The calculated Q-factors are depicted in Fig.~\ref{fig:dipole_c0p27_cond0_Q} for the relative permittivity $\epsilonr=1-\ju 0.25/\omega$. All Q-factors are small for low frequencies where the loss tangent $0.25/\omega$ is high. The Q-factors agree well for approximately  $\ell/\lambda \leq 0.5$ or $k_0a\leq \pi/2$, where $a$ denotes the radius of the smallest circumscribing sphere and $k_0$ is the free-space wavenumber. The differences between the Q-factors from the differentiated impedance $\QabsZp$ in~\eqref{eq:QZp} and reactance $\QXp$ in~\eqref{eq:QXp} matrices are small indicating that the contribution from $\Rm'$ is negligible for this case. The Q-factor from the Brune circuit $\QZB$ follow $\QXp$ well but gives slightly lower values. The Q-factor from the differentiated input impedance is also similar to $\QZB$ except for $\ell/\lambda\approx 2$, where $\QZp$ has a dip, see also~\cite{Gustafsson+Jonsson2015a} for the corresponding free space case where $\QZp\approx 0$ at $\ell/\lambda\approx 2$.
 
\begin{figure}[t]
\begin{center}
\noindent
  \includegraphics[width=0.68\linewidth]{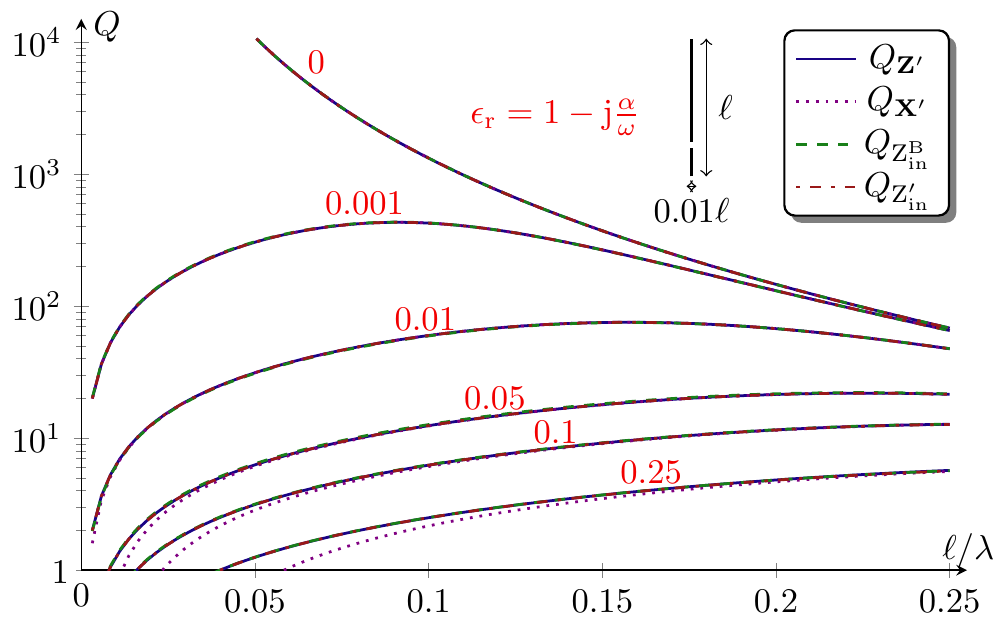}
  \caption{Q-factors for a strip dipole with length $\ell$, width $\ell/100$, fed $0.27\ell$ from the center, and placed in a homogeneous medium with relative permittivity $\epsilonr=1-\alpha\ju/\omega$, with $\alpha=\{0.25,0.1,0.05,0.01,0.001,0\}$ and $\omega=2\pi\ell/\lambda$.}
  \label{fig:dipole_c0p27_cond1_Q}
\end{center}
\end{figure}

The conductivity is swept to produce the relative permittivity $\epsilonr=1-\alpha\ju/\omega$, with $\alpha=\{0.25,0.1,0.05,0.01,0.001,0\}$ for the same strip dipole in Fig.~\ref{fig:dipole_c0p27_cond1_Q}. Here, the differences between the expressions for the Q-factors are negligible. The effect of the temporal dispersion on the differentiated input impedance is estimated by the factor 
\begin{equation}
  \left|\frac{\omega\partial\epsilon}{2\epsilon\partial\omega}\right|
    =\frac{1}{2}\left|\frac{-1}{\frac{\ju\omega}{\alpha}+1}\right|
    =\frac{1/2}{\sqrt{\frac{\omega^2}{\alpha^2}+1}}
    \leq 
    \begin{cases}
      \frac{1}{2} & \text{all }\omega\\
      \frac{1}{\sqrt{8}} & \omega>\alpha
    \end{cases}
\label{eq:}
\end{equation}
indicating that the effect of the dispersion for $\QXp$ is small if $\omega\gg\alpha$. This is also seen in Fig.~\ref{fig:dipole_c0p27_cond1_Q}, where all estimates agree except for $\omega<\alpha$ where also $Q$ is very low.

\subsection{Debye model}
\begin{figure}[t]
\begin{center}
\noindent
  \includegraphics[width=0.68\linewidth]{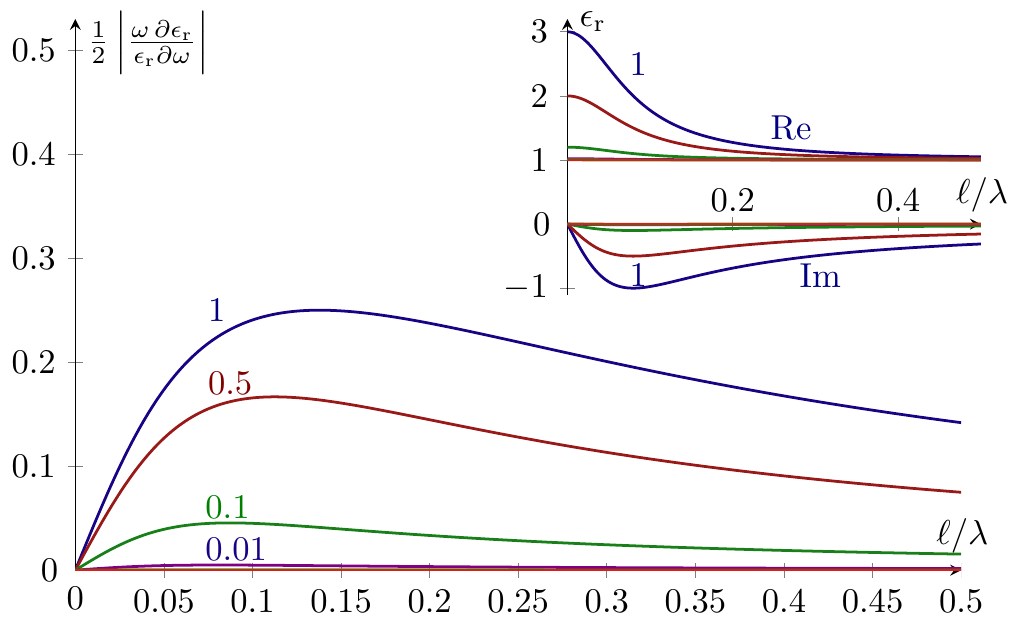}
  \caption{Weighted frequency derivative of the relative permittivity for the Debye model $\epsilonr=1+\frac{\alpha}{0.5+\ju\omega}$ with $\alpha=\{1,0.5,0.1,0.01,0.001,0\}$ and $\omega=2\pi\ell/\lambda$.}
  \label{fig:Loop_Debye_epsp}
\end{center}
\end{figure}

The Debye model describes relaxation effects in molecules and is used to model the permittivity of distilled water and other polar liquids~\cite{Jackson1999}.
The Debye model can be written $\epsilonr=\epsiloninf+(\epsilons-\epsiloninf)/(1+\ju\omega\tau)$, where $\tau$ is the relaxation time, $\epsilons$ the static relative permittivity, and $\epsiloninf$ the high-frequency response. 
Here, we consider the Debye models
\begin{equation*}
  \epsilonr=1+\frac{\alpha}{0.5+\ju\omega}
\label{eq:}
\end{equation*}
giving a dimensionless relaxation time $\tau=2$. The permittivity and derivative $|\omega\epsilonr'/\epsilonr|/2$ in~\eqref{eq:EFIEp} are depicted in Fig.~\ref{fig:Loop_Debye_epsp} for the parameter values $\alpha=\{1,0.5,0.1,0.01,0.001,0\}$. We observe that $|\omega\epsilonr'/\epsilonr|/2\leq 1/4$ that extends to all Debye models.  
 
\begin{figure}[t]
\begin{center}
\noindent
  \includegraphics[width=0.68\linewidth]{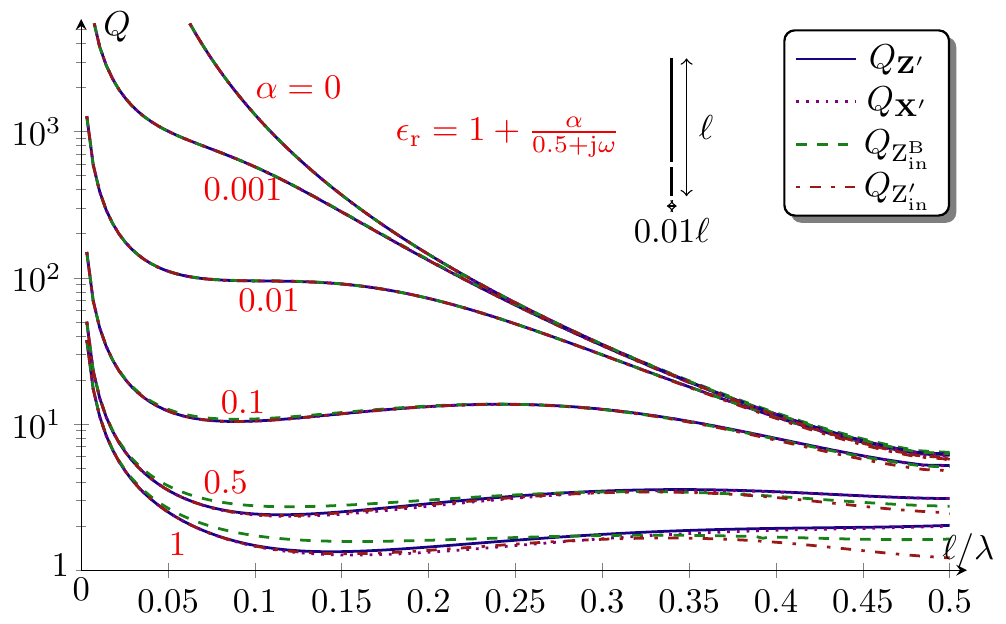}
  \caption{Q-factors for a strip dipole with length $\ell$, width $\ell/100$, fed $0.27\ell$ from the center, and placed in a homogeneous Debye medium with relative permittivity $\epsilonr=1+\frac{\alpha}{0.5+\ju\omega}$ with $\alpha=\{1,0.5,0.1,0.01,0.001,0\}$, see Fig.~\ref{fig:Loop_Debye_epsp}.}
  \label{fig:dipole_c0p27_Debye_Q}
\end{center}
\end{figure}

\begin{figure}[t]
\begin{center}
\noindent
  \includegraphics[width=0.68\linewidth]{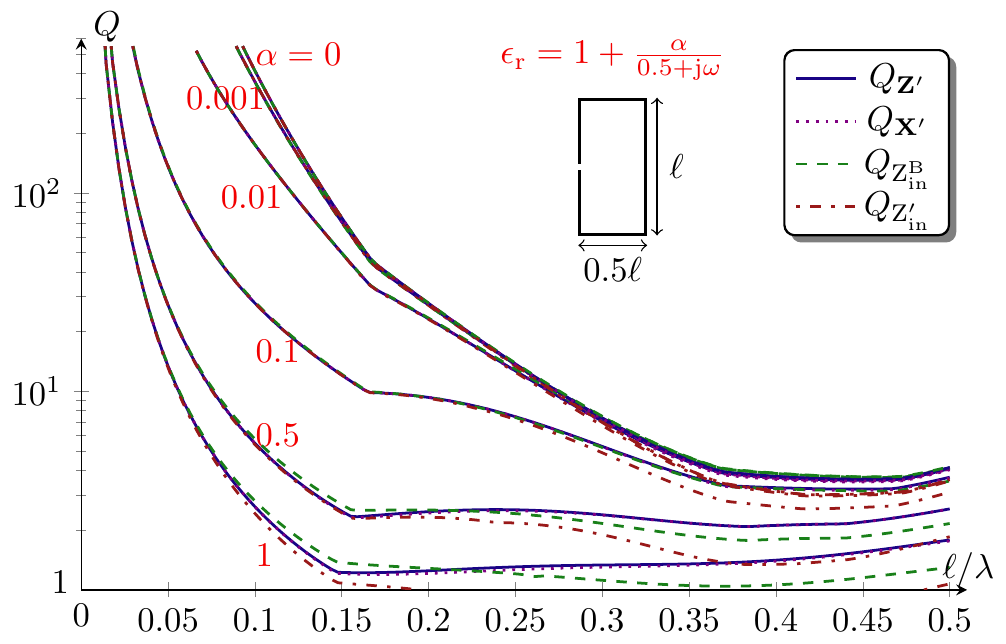}
  \caption{Q-factors for a loop antenna with length $\ell$, width $\ell/2$, fed at the center of the longer side, and placed in a homogeneous medium with relative permittivity $\epsilonr=1+\frac{\alpha}{0.5+\ju\omega}$ with $\alpha=\{1,0.5,0.1,0.01,0.001,0\}$, see Fig.~\ref{fig:Loop_Debye_epsp}.}
  \label{fig:loop_0p5_Debye2_Q}
\end{center}
\end{figure}

The Q-factors from the differentiated input impedance $\QZp$, the Brune synthesized circuit model $\QZB$, the differentiated impedance matrix $\QabsZp$, and the differentiated reactance matrix $\QXp$ are depicted in Figs.~\ref{fig:dipole_c0p27_Debye_Q} and~\ref{fig:loop_0p5_Debye2_Q} for the cases of an off-center fed dipole and a loop antenna, respectively.
We observe the general trend that the Q-factors decrease with increasing $\alpha$ as a result of the increasing losses, see Fig.~\ref{fig:Loop_Debye_epsp}. Moreover, the Q-factors agree well for $Q>5$ and $\QZp$ is slightly below the other curves for $Q<5$. The good agreement is partly explained by the relative small values of $|\omega\epsilonr'/\epsilonr|/2$, as depicted in Fig.~\ref{fig:Loop_Debye_epsp}, indicating that~\eqref{eq:EFIEp} is dominated by the matrix elements in~\eqref{eq:EFIEkp}.

\subsection{Lorentz model}
\begin{figure}[t]
\begin{center}
\noindent
  \includegraphics[width=0.68\linewidth]{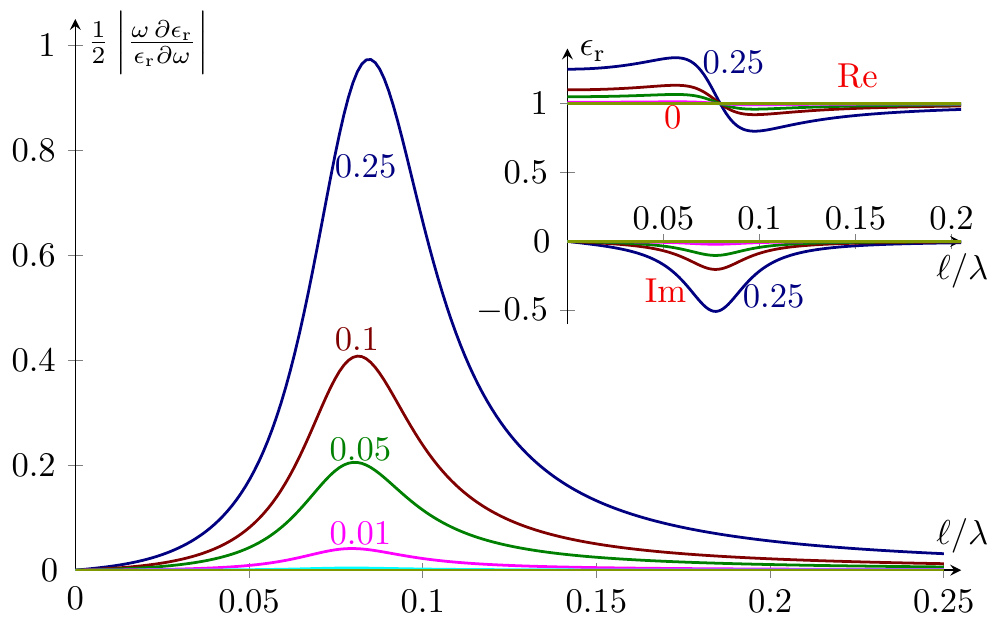}
  \caption{Weighted frequency derivative of the relative permittivity and relative permittivity for the Lorentz model $\epsilonr=1+\frac{\alpha}{1+\ju\omega-4\omega^2}$ with $\alpha=\{0.25,0.1,0.05,0.01,0.001,0\}$ and $\omega=2\pi\ell/\lambda$.}
  \label{fig:dipole_c0p27_Lorentz1_epsp}
\end{center}
\end{figure}

\begin{figure}[t]
\begin{center}
\noindent
  \includegraphics[width=0.68\linewidth]{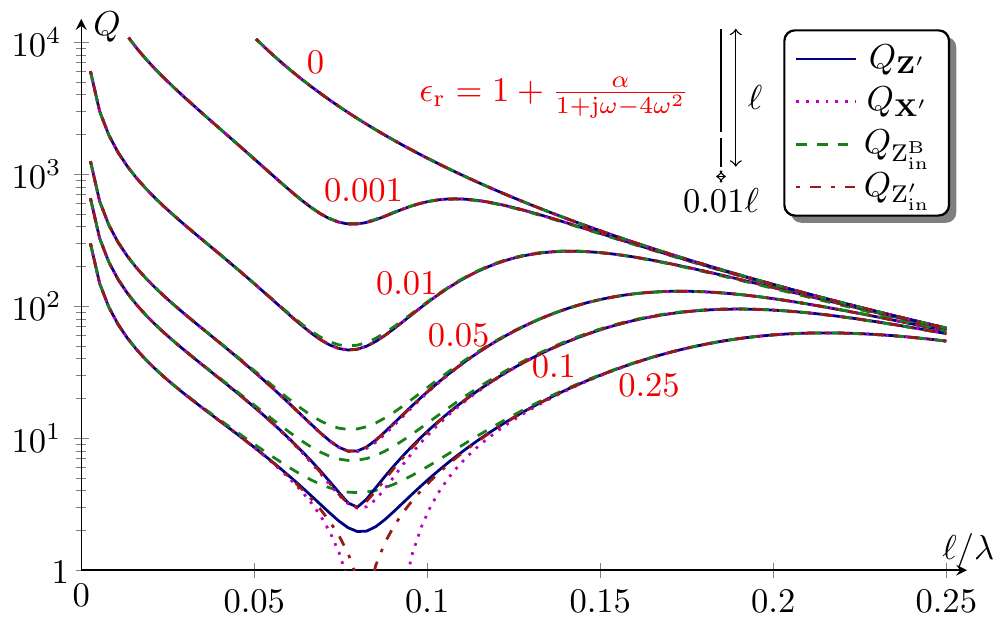}
  \caption{Q-factors for a strip dipole with length $\ell$, width $\ell/100$, fed $0.27\ell$ from the center, and placed in a homogeneous medium with relative permittivity $\epsilonr=1+\frac{\alpha}{1+\ju\omega-4\omega^2}$, where $\omega=2\pi\ell/\lambda$ and $\alpha=\{0.25,0.1,0.05,0.01,0.001,0\}$.}
  \label{fig:dipole_c0p27_Lorentz1_Q}
\end{center}
\end{figure}

The Lorentz model 
\begin{equation}
  \epsilonr = \epsiloninf + \frac{\alpha}{\omega_0^2+\ju\nu\omega\omega_0-\omega^2}
\label{eq:}
\end{equation}
or, more generally, a sum of Lorentz terms are used to describe resonance phenomena in media~\cite{Jackson1999}. We consider Lorentz models of the form
$\epsilonr=1+\frac{\alpha}{1+\ju\omega-4\omega^2}$
with $\alpha=\{0.25,0.1,0.05,0.01,0.001,0\}$
having the resonance frequency $\omega_0=1/2$ or equivalently $\ell/\lambda=1/(4\pi)\approx 0.08$. The permittivity and weighted frequency derivative $|\omega\epsilonr'/\epsilonr|/2$ are depicted in Fig.~\ref{fig:dipole_c0p27_Lorentz1_epsp}. Here, we observe that the weighted frequency derivative is close to unity for the $\alpha=0.25$ case. The matrix elements from~\eqref{eq:EFIEkp} are hence multiplied by an almost arbitrary phase. This implies that the imaginary part of the impedance matrix can assume any value and that the Q-factors defined from the differentiated reactance matrix can be erroneous. This is also observed in Fig.~\ref{fig:dipole_c0p27_Lorentz1_Q}, where the Q-factors for an off-center fed dipole in a background Lorentz media are depicted. The Q-factors agree well for approximately $Q>10$ and start to deviate for lower Q-values. In particular the estimated Q-factors from the differentiated reactance matrix $\QImZp$ and differentiated input impedance $\QZp$ are very low around the resonance frequency $\ell/\lambda\approx 0.08$ for the $\alpha=0.25$ case. 

\subsection{Drude model}
\begin{figure}[t]
\begin{center}
\noindent
  \includegraphics[width=0.68\linewidth]{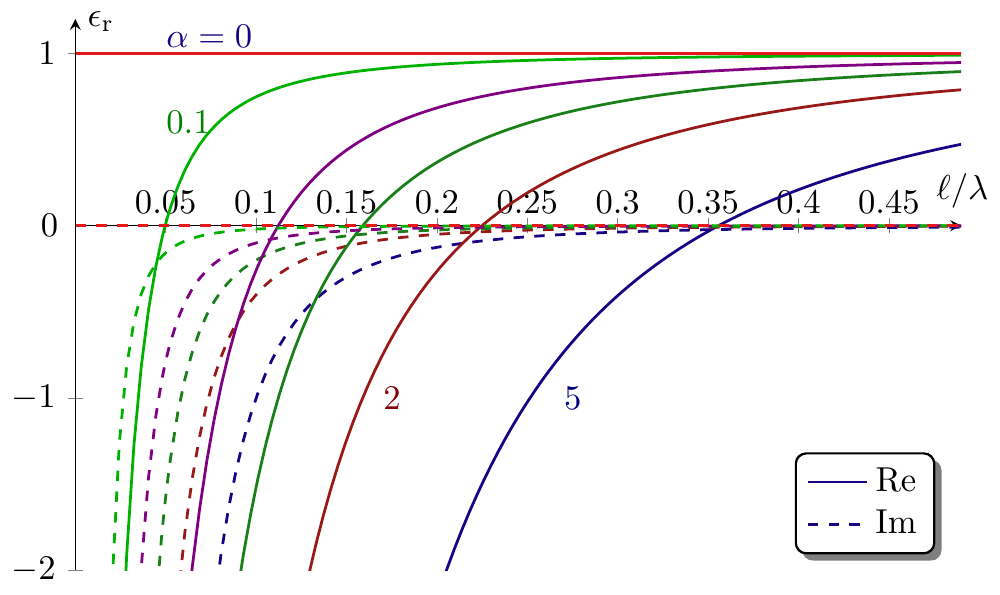}
  \caption{Relative permittivity for the Drude model $\epsilonr=1+\frac{\alpha}{0.05\ju\omega-\omega^2}$ with $\alpha=\{5,2,1,0.5,0.1,0\}$ and $\omega=2\pi\ell/\lambda$.}
  \label{fig:Dipole_Drude_eps}
\end{center}
\end{figure}

The interaction between a free electron gas and electromagnetic fields can be modeled with a Drude dispersion model~\cite{Jackson1999}. The model also exhibits phenomena such as negative permittivity and epsilon near zero materials~\cite{Alu+etal2007}. Consider the relative permittivity 
\begin{equation}
  \epsilonr = 1+\frac{\alpha}{0.05\ju\omega-\omega^2}
\label{eq:Drudemodel}
\end{equation}
as depicted in Fig.~\ref{fig:Dipole_Drude_eps} for the parameter values $\alpha=\{5,2,1,0.5,0.1,0\}$. We observe that the real part of the permittivity is negative for low frequencies and that the permittivity is close to zero around $\omega=\sqrt{\alpha}$ or $\ell/\lambda\approx 0.16\sqrt{\alpha}$.

\begin{figure}[t]
\begin{center}
\noindent
  \includegraphics[width=0.68\linewidth]{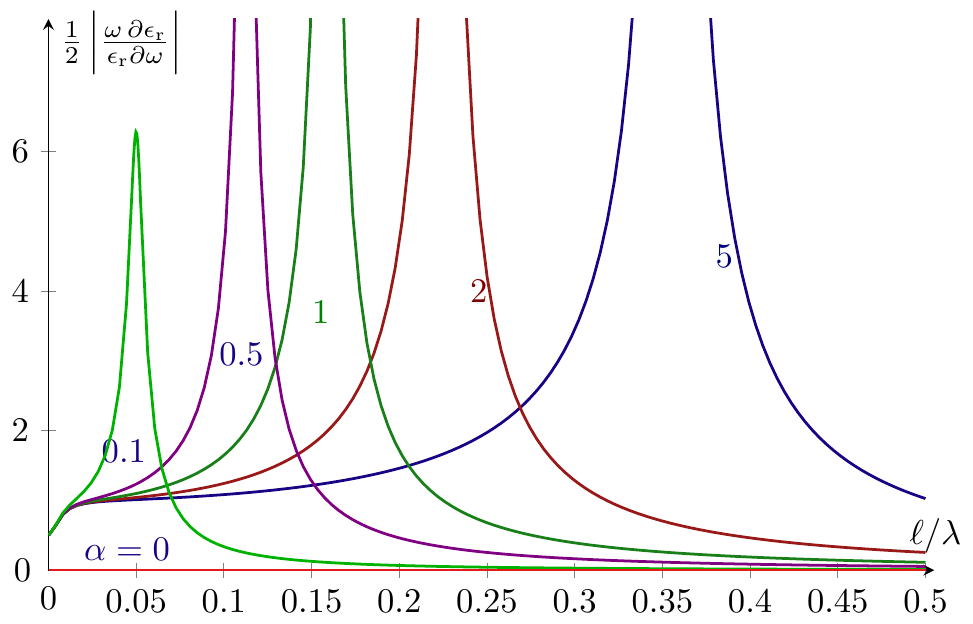}
  \caption{Weighted frequency derivative of the relative permittivity for the Drude models depicted in Fig.~\ref{fig:Dipole_Drude_eps}.}
  \label{fig:Dipole_Drude1_epsp}
\end{center}
\end{figure}

The weighted frequency derivative of the Drude model~\eqref{eq:Drudemodel} is shown in Fig.~\ref{fig:Dipole_Drude1_epsp}. We observe that the weighted frequency derivative is maximal for the frequency $\omega=2\pi\ell/\lambda$ where the permittivity is close to zero. The maximal value is also much greater than unity indicating that the matrix elements~\eqref{eq:EFIEkp} are multiplied with a complex number of arbitrary phase. 

\begin{figure}[t]
\begin{center}
\noindent
  \includegraphics[width=0.68\linewidth]{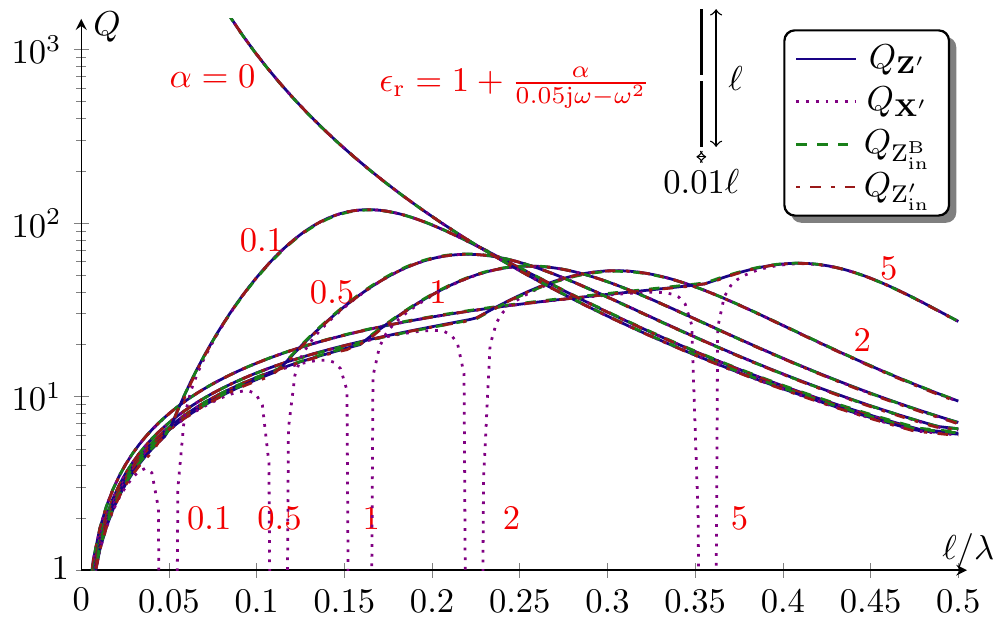}
  \caption{Q-factors for a strip dipole with length $\ell$, width $\ell/100$, fed at the center, and placed in a homogeneous Drude medium with relative permittivity $\epsilonr=1+\frac{\alpha}{0.05\ju\omega-\omega^2}$, where $\omega=2\pi\ell/\lambda$ and $\alpha=\{5,2,1,0.5,0.1,0\}$, see Figs~\ref{fig:Dipole_Drude_eps} and~\ref{fig:Dipole_Drude1_epsp}.}
  \label{fig:dipole_0p01_cf_Drude_Q}
\end{center}
\end{figure}

The resulting Q-factors for a center fed dipole embedded in a homogeneous Drude model are shown in Fig.~\ref{fig:dipole_0p01_cf_Drude_Q}. The Q factors determined from the Brune synthesized circuit $\QZB$, the differentiated input impedance $\QZp$, and differentiated impedance matrix $\QabsZp$ agree very well for all considered cases. The Q-factors from the differentiated reactance matrix $\QImZp$ agree with the other estimates except for frequencies where the weighted frequency derivative of the relative permittivity is large. This is consistent with the interpretation of the multiplication of the matrix element in~\eqref{eq:EFIEkp} with a large complex valued number giving the resultant matrix element an arbitrary phase and hence potentially a small imaginary part.

In Fig.~\ref{fig:dipole_0p01_cf_Drude_Q}, we also observe that the Q-factors decrease for low frequencies where the losses are high and the real part of the permittivity is negative, \cf Fig.~\ref{fig:Dipole_Drude_eps}. The Q-factors are higher than for the free-space case ($\alpha=0$) in the region where the losses are small and the real part of the relative permittivity is between zero and unity. This is partly explained by a comparison with the case of a dipole in a homogeneous non-dispersive lossless medium with a relative permittivity in the range $0<\epsilonr<1$. Here, the lower permittivity corresponds to an increased wavelength or equivalently a shorter dipole. Consider \eg the $\alpha=5$ case that has the relative permittivity $\epsilonr\approx 0.5$ at $\ell/\lambda=0.5$. This corresponds to the free space case at $\sqrt{0.5}\ 0.5\approx 0.35$ with a Q-factor approximately 20. We also observe that although there is a small change in the Q-factors at the frequencies where $\epsilonr\approx 0$, there is not a major effect of $\epsilonr\approx 0$, except for the approximation using $\QImZp$.

\subsection{Refractive index}

We synthesize a medium such that $\Zm'\approx \mat{0}$ at a desired frequency $\omega_0$. This implies that $\QZp\approx 0$ for any self-resonant antenna at $\omega_0$ unless $P_{\mrm{d}}=\Jm^{\herm}\Rm\Jm/2\approx 0$ at the same frequency. It is known that one can synthesize antennas with $\QZp\approx 0$ as shown in~\cite{Gustafsson+Nordebo2006b} and also by the construction in~\eqref{eq:Zmp_Takagi}. However, the explicit construction used here also shows that it is not possible to express the stored energy solely in $\Zm'$ that is valid for general (passive) temporal dispersive media. This implies that although the considered energy expressions~\eqref{eq:WemJV}, \eqref{eq:QZp}, and~\eqref{eq:QYp} work well for the considered dispersion models they do not work for every dispersion model.

For simplicity we consider models with identical relative permittivity and permeability, $\epsilonr=\mur$, and hence $\eta=\eta_0$ and $k=\epsilonr\omega/\clight$. 
The term
\begin{equation}
  \frac{\omega}{k}\partder{k}{\omega} 
  =\frac{1}{\epsilonr}\partder{\omega\epsilonr}{\omega}
\label{eq:label:LorentzN_dwdk}
\end{equation}
multiplies the matrix elements in~\eqref{eq:EFIEp1}. This term vanishes if $(\omega\epsilon)'=0$ unless $\epsilon=0$ simultaneously. The properties of $(\omega\epsilon)'$ are well understood and classical results~\cite{Landau+Lifshitz1960,Yaghjian2007,Gustafsson+Sjoberg2010a} show that $(\omega\epsilon)'\geq\epsiloninf$ in frequency intervals with $\Im\epsilonr=0$, where $\epsiloninf=\lim_{\omega\to\infty}\epsilonr/\omega$ is the high-frequency limit of the permittivity, that is often assumed equal to unity $\epsiloninf=1$,~\cite{Landau+Lifshitz1960}. This suggests that $(\omega\epsilon)'$ does not vanish in lossless media. However, the assumption of a lossless medium is essential for the bounds presented in~\cite{Landau+Lifshitz1960,Yaghjian2007}. The bounds are generalized to lossy media in~\cite{Gustafsson+Sjoberg2010a}, where it is demonstrated that there are no point wise (at a single frequency) bounds on $(\omega\epsilon)'$ for general passive material models, see also App.~\ref{S:CircuitDisp}.

\begin{figure}[t]
\begin{center}
\noindent
  \includegraphics[width=0.68\linewidth]{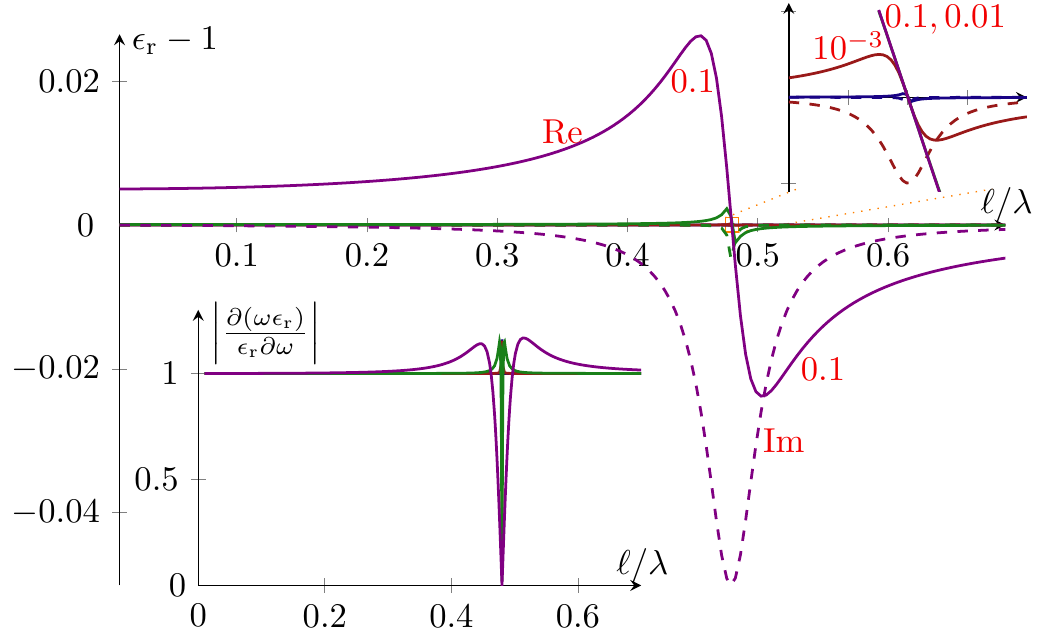}
  \caption{Lorentz resonance relative permittivity and permeability in~\eqref{eq:LorentzN} with $\omega=2\pi\ell/\lambda$, $\omega_0=0.48\ 2\pi$, and $\nu=10^{-n}$ with $n=\{1,2,3,4\}$. A zoomed in part around the resonance $\omega_0$ is depicted in the top right inset. The magnitude of the factor~\eqref{eq:label:LorentzN_dwdk} is depicted in the lower left inset.}
  \label{fig:sLorentz_eps}
\end{center}
\end{figure}

Consider the Lorentz model~\cite{Gustafsson+Sjoberg2010a}
\begin{equation}
  \epsilonr=\mur=1+\frac{\nu^{2}\omega_0^2/2}{\omega_0^2-\omega^2+\ju\omega\nu\omega_0},
\label{eq:LorentzN}
\end{equation}
see Fig.~\ref{fig:sLorentz_eps}, having the values
\begin{equation}
  \epsilonr(\omega_0) = 1-\ju\nu/4
  \qtext{and }
  \left.(\omega\epsilonr)'\right|_{\omega=\omega_0} = 0
\label{eq:}
\end{equation}
at the resonance frequency and implying that $\Zm'=\mat{0}$ at $\omega=\omega_0$ for any $\nu> 0$, \cf~\eqref{eq:EFIEp1}. The corresponding impedance matrix does not change significantly as $\nu\to 0$ and hence does not the energy distribution in the fields, currents, or circuit models of the antenna. 

\begin{figure}[t]
\begin{center}
\noindent
  \includegraphics[width=0.68\linewidth]{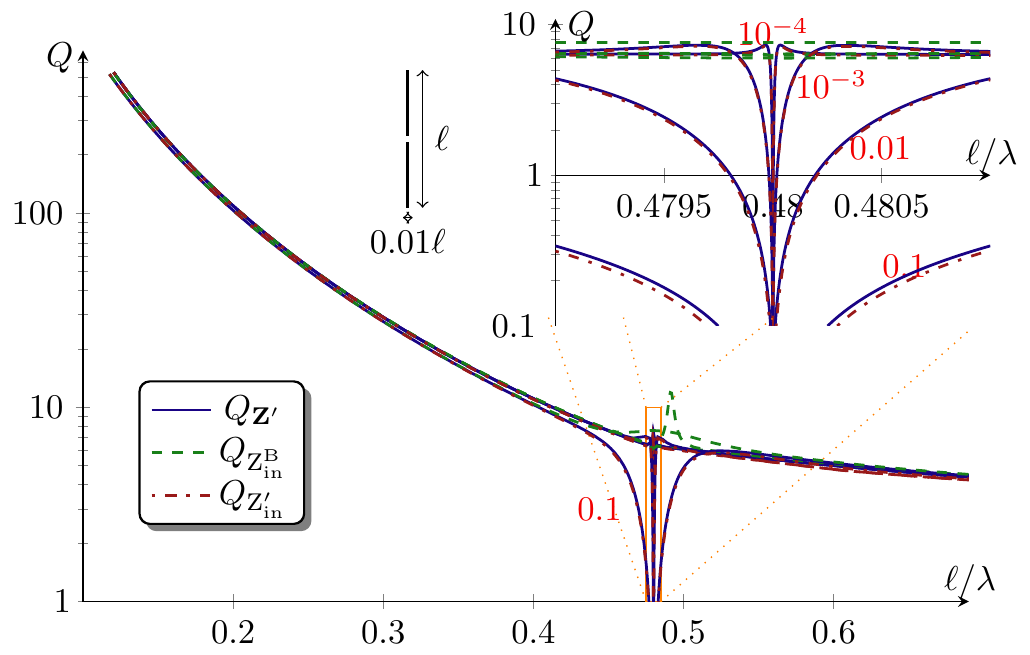}
  \caption{Q-factors for a strip dipole with length $\ell$, width $\ell/100$, fed at the center, and placed in a homogeneous electric and magnetic Lorentz medium with relative permittivity and permeability~\eqref{eq:LorentzN} as depicted in Fig.~\ref{fig:sLorentz_eps}, where $\omega=2\pi\ell/\lambda$ and $\nu=10^{-n}$ with $n=\{1,2,3,4\}$. The region around the resonance frequency is depicted in the upper right inset.}
  \label{fig:dipole_0p01_cf_sLorentz_Q}
\end{center}
\end{figure}

The computed Q-factor values for a strip dipole with length $\ell$ and width $0.01\ell$ are used to illustrate the estimated Q-factors, see Fig.~\ref{fig:dipole_0p01_cf_sLorentz_Q}. Consider the resonance frequency $\omega_0=2\pi\ell/\lambda=3$ and the damping $\nu=10^{-n}$ for $n=0,1,2,3$ in the Lorentz model~\eqref{eq:LorentzN}, see Fig.~\ref{fig:sLorentz_eps}. The maximal susceptibility is $|1-\epsilonr|=\nu/\sqrt{4-\nu^2}\approx \nu/2$ for $\nu\ll 1$. The magnitude of the frequency derivative $\frac{\omega}{k}\partder{k}{\omega}=\epsilonr^{-1}\partder{(\omega\epsilonr)}{\omega}$ is depicted in Fig.~\ref{fig:sLorentz_eps}. Here, it is observed that $|\epsilonr^{-1}(\omega\epsilonr)'|$ is zero at the resonance frequency $\omega=\omega_0$ and close to unity away from the resonance. The resonance has a relative half-power bandwidth proportional to $\nu$. 

\section*{Conclusions}
The analysis of the frequency derivative of the input impedance and the stored electromagnetic energy are unified by the frequency derivative of the electric field integral equation (EFIE) impedance matrix $\Zm'$. The differentiated input impedance is proportional to the bilinear quadratic form $\Jm^{\tran}\Zm'\Jm$. The corresponding stored energy for lumped circuit networks is the sesquilinear quadratic form $\Jm^{\herm}\Zm'\Jm/4$. This shows that they basically differ by a complex conjugate of the current matrix. The resulting Q-factors are hence similar for currents with negligible phase variation. 

The stored energy for electromagnetic systems is more involved. Here, we show that the stored energy introduced by Vandenbosch~\cite{Vandenbosch2010} in the free space case is identical to the quadratic form $\Jm^{\herm}\Xm'\Jm/4$ of the differentiated reactance matrix, see also~\cite{Harrington1975}. This energy expression has been verified for several antennas with good results~\cite{Hazdra+etal2011,Gustafsson+Jonsson2015a,Cismasu+Gustafsson2014a,Cismasu+Gustafsson2014b}. In~\cite{Gustafsson+etal2012a}, it is however shown that the quadratic form can be indefinite for sufficiently large structures. This partly questions the validity of the energy expression, although the same problem appears in the commonly used stored energy~\cite{Yaghjian+Best2005} defined by subtraction of the far-field~\cite{Gustafsson+Jonsson2014}.  
  
We investigate possible generalizations of the stored energy based on the quadratic form  $|\Jm^{\herm}\Zm'\Jm|/4$. These expressions resemble the $\QZp$ formula based on $|\Jm^{\tran}\Zm'\Jm|$. The proposed expressions are illustrated for dipole and loop antennas integrated in temporally dispersive media. The resulting Q-factors are compared with the Q-factor determined from Brune synthesized circuit networks. The results suggest that the proposed expressions are valid as long as the losses are not too large and the temporal dispersion (frequency dependence) is not too strong. We also synthesize a temporally dispersive material model $(\omega\epsilon)'=0$ that gives $\Zm'=\mat{0}$, hence vanishing $\QZp$ and $\QabsZp$ values. This shows that the proposed stored energy expressions are not valid for all material models. This is also the case for the classical definition of energy density, \ie $(\omega\epsilon)'|\Ev|^2/4+(\omega\mu)'|\Hv|^2/4$ in dispersive media~\cite{Landau+Lifshitz1960}.

The proposed quadratic form for the differentiated input impedance $\Zin'$ can also be useful for efficient interpolation of $\Zin$ over a frequency interval. In essence, the additional computational cost to evaluate $\Zin'$ is small compared to the evaluation of $\Zin$ as $\Jm$ is already computed. This means that the interpolation can be performed with both $\Zin$ and $\Zin'$ known in a set of frequency points. The same procedure can also be used to 
evaluate higher order derivatives. The technique could potentially be used to evaluate the frequency derivatives of other parameters.

The used approach can lastly be used to generalize the results to more complex material models such as anisotropic and bi-anisotropic material models, as well as evaluation of $\QZp$ for inhomogeneous structures.

\appendix
\section{Circuit models for temporal dispersion}\label{S:CircuitDisp}
The temporal dispersion of background media enters the energy expressions in the multiplicative terms involving the classical frequency derivatives $(\omega\epsilon)'$ in~\eqref{eq:EFIEp1}.  
The approximation of the electric energy density as the real or absolute value of $(\omega\epsilon)'|\Ev|^2/4$ is accurate for many material models but has obvious problems with material models such as, the resonance model~\eqref{eq:LorentzN} where $(\omega\epsilon)'=0$. This resembles the antenna case with $\QZinp=0$ although $Q\gg 1$ in~\cite{Gustafsson+Nordebo2006b}. One possible solution is to use a circuit model for the temporal dispersion and to determine the corresponding total stored energy in the circuit model, see also~\cite{Ruppin2002,Tretyakov2005}. This is similar to calculate the antenna Q from a Brune synthesized circuit model~\cite{Gustafsson+Jonsson2015a}. 

Consider for simplicity the case with a general Lorentz type resonance, \ie
\begin{equation}
  \epsilonr(\omega) = \epsiloninf + \frac{\alpha}{\beta+\ju\gamma\omega-\delta\omega^2}
\label{eq:}
\end{equation}
that includes the conductivity, Debye, and Drude models as special cases. The corresponding time-domain representation expresses the electric flux density $\Dv$ as
\begin{equation}
  \Dv = \epsilon_0\epsiloninf\Ev + \Pv
\label{eq:}
\end{equation} 
where the polarization $\Pv$ satisfies the ordinary differential equation
\begin{equation}
  \delta\ddot{\Pv} + \gamma\dot{\Pv} + \beta\Pv 
  = \alpha\epsilon_0\Ev 
\label{eq:Pdiff}
\end{equation}
and the dot denotes differentiation with respect to time.
Insertion of $\Dv$ into Maxwell's equations and multiplication with $\Ev$ and $\dot{\Pv}$ gives the energy balance
\begin{equation}
  \Ev\cdot\dot{\Dv} = \frac{\epsilon_0}{2}\partder{}{t}
  \big(\epsiloninf |\Ev|^2 + \frac{\delta|\dot{\Pv}|^2}{\alpha\epsilon_0^2}
  +\frac{\beta|\Pv|^2}{\alpha\epsilon_0^2}\big) 
  +\frac{\gamma|\dot{\Pv}|}{\alpha\epsilon_0},
\label{eq:TimeSenergy}
\end{equation}
where we identify the terms differentiated with respect to time as the terms contributing to the  stored energy. 

Transforming back to the frequency domain gives the electric energy density~\cite{Ruppin2002,Tretyakov2005}
\begin{equation}
  w_{\mrm{e}}=\frac{\epsilon_0}{4}
  \big(\epsiloninf |\Ev|^2 + \frac{\delta\omega^2
  +\beta}{\alpha\epsilon_0^2}|\Pv|^2\big)
  =\frac{\epsilon_0}{4}
  \left(\epsiloninf  + \frac{\alpha(\delta\omega^2
  +\beta)}{(\beta-\delta\omega^2)^2+\omega^2\gamma^2}\right)|\Ev|^2.  
\label{eq:WeLorentz}
\end{equation} 
The special case of a conductivity model ($\beta=\delta=0$) reduces to the classical approximation $w_{\mrm{e}}=(\omega\epsilon)'|\Ev|^2/4=\epsilon_0\epsiloninf|\Ev|^2/4$, but Debye, Drude, and Lorentz models can give different results. 

\begin{figure}[t]
\begin{center}
\noindent
  \includegraphics[width=0.48\linewidth]{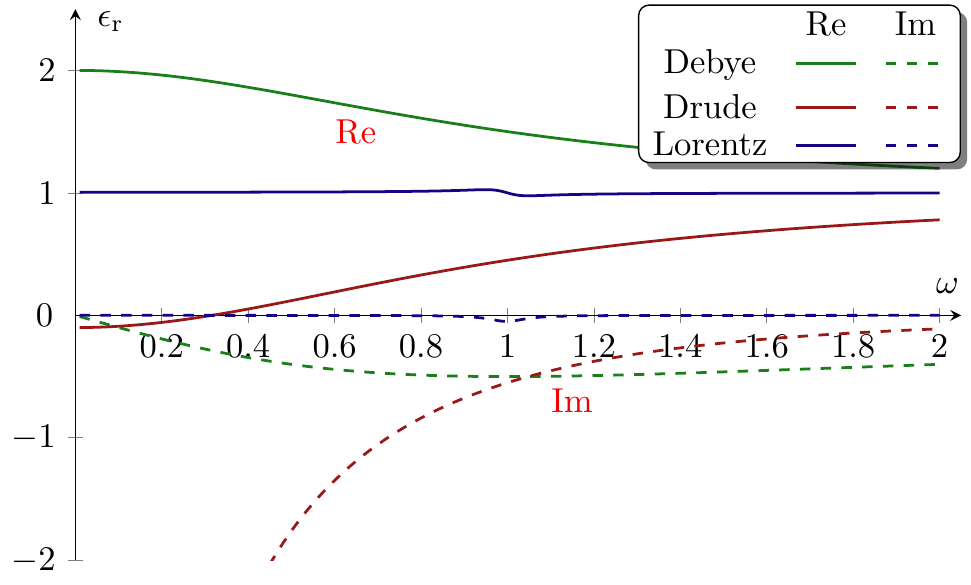}
  \includegraphics[width=0.48\linewidth]{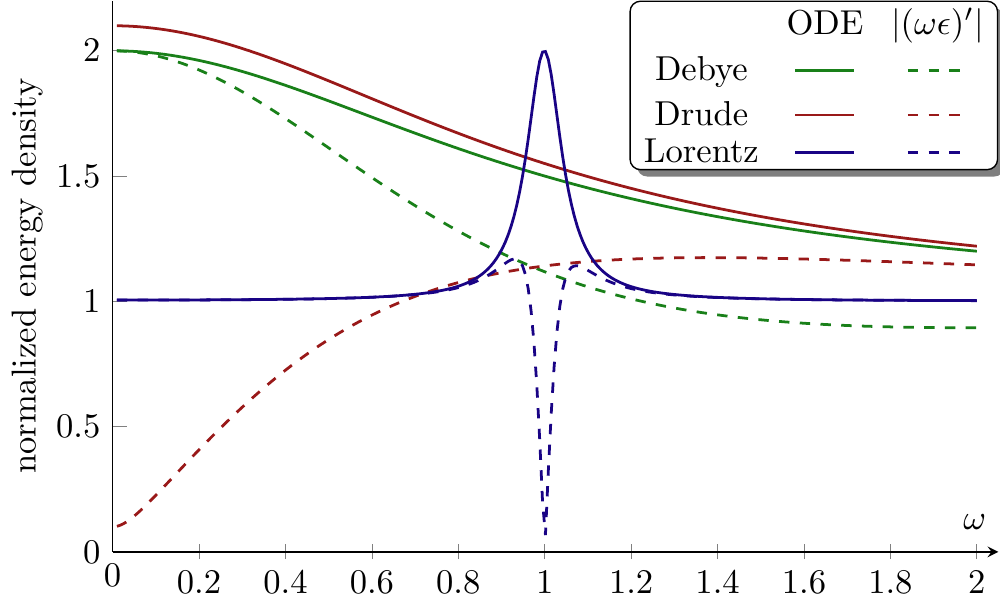}
  \caption{Relative permittivity and normalized energy density for the Debye model $\epsilonr=1+1/(1+s)$, Drude model $\epsilonr=1+1.1/(s+s^2)$, and Lorentz model $\epsilonr=1+0.005/(1+0.1s+s^2)$, with $s=\ju\omega$. The normalized energy density $w_{\mrm{e}}/(\epsilon_0|\Ev|^2/4)$ in solid curves and local approximation $|(\omega\epsilon)'|$ in dashed curves.}
  \label{fig:EnergyDensityDDL}
\end{center}
\end{figure}

The Debye model $\epsilonr=1+1/(1+s)$, Drude model $\epsilonr=1+1.1/(s+s^2)$, and Lorentz model $\epsilonr=1+0.005/(1+0.1s+s^2)$ with $s=\ju\omega$ as depicted in Fig.~\ref{fig:EnergyDensityDDL} are used to illustrate the differences between the normalized energy density $w_{\mrm{e}}4/|\Ev|^2$ from~\eqref{eq:WeLorentz} and the local approximation $|(\omega\epsilon)'|$. We note that the $w_{\mrm{e}}$ values are similar for the Debye and Drude models. This is explained by the decomposition of the Drude model as the difference between a conductivity term and a Debye term, \ie $1/(s+s^2)=1/s-1/(1+s)$, and that the conductivity term does not contribute to the energy density. The corresponding local approximation $|(\omega\epsilon)'|$ differ substantially for the Debye and Drude cases due to the sign change in the Debye term. The Lorentz term has $(\omega\epsilon)'=0$ for $\omega=1$. The normalized energy density~\eqref{eq:WeLorentz} increases to 2 for $\omega=1$.

The relation between the energy density determined from the local approximation $(\omega\epsilonr)'$ and the differential equation~\eqref{eq:Pdiff} is similar to the relation between the Q-factors defined by differentiation of the input impedance $\QZinp$ and circuit synthesis $\QZB$. The analogy is clearly seen by identification of the input impedance as $Z_{\epsilon}=s\epsilon$. 
We synthesize a circuit model from the equivalent input impedance 
\begin{equation}
  Z_{\epsilon}(s) = s\epsilonr(s)
  =s\epsiloninf+\frac{\alpha s}{\beta+\gamma s+\delta s^2}
\label{eq:}
\end{equation}
that has the circuit representation
\begin{center}
\begin{circuitikz} \draw
(0,2) node[anchor=east]{}
to[L, l_=$\epsiloninf$, *-] (2,2)
to[short] (5,2)
to[R, l_=$\displaystyle\frac{\alpha}{\gamma}$] (5,0)
to[short, -*] (0,0)
(2,2) to[L, l_=$\displaystyle\frac{\alpha}{\beta}$, *-*] (2,0)
(3.5,2) to[C, l_=$\displaystyle\frac{\delta}{\alpha}$, *-*] (3.5,0)
;\end{circuitikz}
\end{center}
The circuit representation has stored energy in the capacitors and inductors. The stored energy in the capacitors and inductors are usually separated as
electric and magnetic, respectively. There is only electric energy related to the permittivity model~\eqref{eq:Pdiff} so here we consider the total energy in the circuit. The total stored energy in the circuit network is identical to~\eqref{eq:WeLorentz}. Here, we note the Brune synthesis can be used to construct circuit models for arbitrary permittivity models. It is however not clear if the stored energy in the circuit model is identical to the energy density in the electric field.


\begin{thebibliography}{10}
\providecommand{\url}[1]{#1}
\csname url@samestyle\endcsname
\providecommand{\newblock}{\relax}
\providecommand{\bibinfo}[2]{#2}
\providecommand{\BIBentrySTDinterwordspacing}{\spaceskip=0pt\relax}
\providecommand{\BIBentryALTinterwordstretchfactor}{4}
\providecommand{\BIBentryALTinterwordspacing}{\spaceskip=\fontdimen2\font plus
\BIBentryALTinterwordstretchfactor\fontdimen3\font minus
  \fontdimen4\font\relax}
\providecommand{\BIBforeignlanguage}[2]{{%
\expandafter\ifx\csname l@#1\endcsname\relax
\typeout{** WARNING: IEEEtran.bst: No hyphenation pattern has been}%
\typeout{** loaded for the language `#1'. Using the pattern for}%
\typeout{** the default language instead.}%
\else
\language=\csname l@#1\endcsname
\fi
#2}}
\providecommand{\BIBdecl}{\relax}
\BIBdecl

\bibitem{Landau+Lifshitz1960}
L.~D. Landau and E.~M. Lifshitz, \emph{Electrodynamics of Continuous Media},
  1st~ed.\hskip 1em plus 0.5em minus 0.4em\relax Oxford: Pergamon, 1960.

\bibitem{Jackson1999}
J.~D. Jackson, \emph{Classical Electrodynamics}, 3rd~ed.\hskip 1em plus 0.5em
  minus 0.4em\relax New York: John Wiley \& Sons, 1999.

\bibitem{VanBladel2007}
J.~G. Van~Bladel, \emph{Electromagnetic Fields}, 2nd~ed.\hskip 1em plus 0.5em
  minus 0.4em\relax Piscataway, NJ: IEEE Press, 2007.

\bibitem{Engheta+Ziolkowski2006}
N.~Engheta and R.~W. Ziolkowski, \emph{Metamaterials: physics and engineering
  explorations}.\hskip 1em plus 0.5em minus 0.4em\relax John Wiley \& Sons,
  2006.

\bibitem{Capolino2009a}
F.~Capolino, Ed., \emph{Theory and Phenomena of Metamaterials}.\hskip 1em plus
  0.5em minus 0.4em\relax CRC Press, 2009.

\bibitem{Caloz2011}
C.~Caloz, ``Metamaterial dispersion engineering concepts and applications,''
  \emph{Proceedings of the IEEE}, vol.~99, no.~10, pp. 1711--1719, 2011.

\bibitem{Ruppin2002}
R.~Ruppin, ``Electromagnetic energy density in a dispersive and absorptive
  material,'' \emph{Physics letters A}, vol. 299, no.~2, pp. 309--312, 2002.

\bibitem{Tretyakov2005}
S.~Tretyakov, ``Electromagnetic field energy density in artificial microwave
  materials with strong dispersion and loss,'' \emph{Physics Letters A}, vol.
  343, no.~1, pp. 231--237, 2005.

\bibitem{Vorobyev2012}
O.~B. Vorobyev, ``Energy density of macroscopic electric and magnetic fields in
  dispersive medium with losses,'' \emph{Progress In Electromagnetics Research
  B}, vol.~40, pp. 343--360, 2012.

\bibitem{Yaghjian+etal2013}
A.~D. Yaghjian, M.~Gustafsson, and B.~L.~G. Jonsson, ``Minimum {Q} for lossy
  and lossless electrically small dipole antennas,'' \emph{Progress In
  Electromagnetics Research}, vol. 143, pp. 641--673, 2013.

\bibitem{Wheeler1958}
H.~A. Wheeler, ``Fundamental limitations of a small vlf antenna for
  submarines,'' \emph{IRE Trans. on Antennas and Propagation}, vol.~6, pp.
  123--125, 1958.

\bibitem{Skrivervik2013}
A.~K. Skrivervik, ``Implantable antennas: The challenge of efficiency,'' in
  \emph{Antennas and Propagation (EuCAP), 2013 7th European Conference
  on}.\hskip 1em plus 0.5em minus 0.4em\relax IEEE, 2013, pp. 3627--3631.

\bibitem{Moore1963}
R.~K. Moore, ``Effects of a surrounding conducting medium on antenna
  analysis,'' \emph{IEEE Trans. Antennas Propagat.}, vol.~11, no.~3, pp.
  216--225, 1963.

\bibitem{Yaghjian+Best2005}
A.~D. Yaghjian and S.~R. Best, ``Impedance, bandwidth, and {$Q$} of antennas,''
  \emph{IEEE Trans. Antennas Propagat.}, vol.~53, no.~4, pp. 1298--1324, 2005.

\bibitem{Yaghjian2007}
A.~D. Yaghjian, ``Internal energy, {Q}-energy, {P}oynting's theorem, and the
  stress dyadic in dispersive material,'' \emph{IEEE Trans. Antennas
  Propagat.}, vol.~55, no.~6, pp. 1495--1505, 2007.

\bibitem{Hansen+etal2014}
T.~V. Hansen, O.~S. Kim, and O.~Breinbjerg, ``Properties of sub-wavelength
  spherical antennas with arbitrarily lossy magnetodielectric cores approaching
  the {C}hu lower bound,'' \emph{IEEE Trans. Antennas Propagat.}, vol.~62,
  no.~3, pp. 1456--1460, 2014.

\bibitem{Karlsson2004}
A.~Karlsson, ``Physical limitations of antennas in a lossy medium,'' \emph{IEEE
  Trans. Antennas Propagat.}, vol.~52, pp. 2027--2033, 2004.

\bibitem{Harrington1975}
R.~Harrington, ``\BIBforeignlanguage{English}{Characteristic modes for antennas
  and scatterers},'' in \emph{\BIBforeignlanguage{English}{{Numerical and
  Asymptotic Techniques in Electromagnetics}}}, ser. {Topics in Applied
  Physics}, R.~Mittra, Ed.\hskip 1em plus 0.5em minus 0.4em\relax Springer
  Berlin Heidelberg, 1975, vol.~3, pp. 51--87.

\bibitem{Geyi2003b}
W.~Geyi, ``A method for the evaluation of small antenna {Q},'' \emph{IEEE
  Trans. Antennas Propagat.}, vol.~51, no.~8, pp. 2124--2129, 2003.

\bibitem{Vandenbosch2010}
G.~A.~E. Vandenbosch, ``Reactive energies, impedance, and {Q} factor of
  radiating structures,'' \emph{IEEE Trans. Antennas Propagat.}, vol.~58,
  no.~4, pp. 1112--1127, 2010.

\bibitem{Capek+etal2014}
M.~Capek, L.~Jelinek, P.~Hazdra, and J.~Eichler, ``The measurable {Q} factor
  and observable energies of radiating structures,'' \emph{IEEE Trans. Antennas
  Propagat.}, vol.~62, no.~1, pp. 311--318, Jan 2014.

\bibitem{Gustafsson+Jonsson2014}
M.~Gustafsson and B.~L.~G. Jonsson, ``Stored electromagnetic energy and antenna
  {Q},'' \emph{Progress In Electromagnetics Research}, vol. 150, pp. 13--27,
  2014.

\bibitem{Gustafsson+Nordebo2006b}
M.~Gustafsson and S.~Nordebo, ``Bandwidth, {Q} factor, and resonance models of
  antennas,'' \emph{Progress in Electromagnetics Research}, vol.~62, pp. 1--20,
  2006.

\bibitem{Gustafsson+etal2012a}
M.~Gustafsson, M.~Cismasu, and B.~L.~G. Jonsson, ``{Physical bounds and optimal
  currents on antennas},'' \emph{IEEE Trans. Antennas Propagat.}, vol.~60,
  no.~6, pp. 2672--2681, 2012.

\bibitem{Gustafsson+Jonsson2015a}
M.~Gustafsson and B.~L.~G. Jonsson, ``Antenna {Q} and stored energy expressed
  in the fields, currents, and input impedance,'' \emph{IEEE Trans. Antennas
  Propagat.}, 2015, in press.

\bibitem{Hazdra+etal2011}
P.~Hazdra, M.~Capek, and J.~Eichler, ``Radiation {Q}-factors of thin-wire
  dipole arrangements,'' \emph{Antennas and Wireless Propagation Letters,
  IEEE}, vol.~10, pp. 556--560, 2011.

\bibitem{Vandenbosch2013a}
G.~A.~E. Vandenbosch, ``Radiators in time domain, part {I}: electric, magnetic,
  and radiated energies,'' \emph{IEEE Trans. Antennas Propagat.}, vol.~61,
  no.~8, pp. 3995--4003, 2013.

\bibitem{Vandenbosch2013b}
------, ``Radiators in time domain, part {II}: finite pulses, sinusoidal regime
  and {Q} factor,'' \emph{IEEE Trans. Antennas Propagat.}, vol.~61, no.~8, pp.
  4004--4012, 2013.

\bibitem{Wing2008}
O.~Wing, \emph{Classical Circuit Theory}.\hskip 1em plus 0.5em minus
  0.4em\relax New York: Springer, 2008.

\bibitem{Peterson+Ray+Mittra1998}
A.~F. Peterson, S.~L. Ray, and R.~Mittra, \emph{Computational Methods for
  Electromagnetics}.\hskip 1em plus 0.5em minus 0.4em\relax New York: IEEE
  Press, 1998.

\bibitem{Pozar1983}
D.~M. Pozar, ``Considerations for millimeter wave printed antennas,''
  \emph{IEEE Trans. Antennas Propagat.}, vol.~31, no.~5, pp. 740--747, Sep.
  1983.

\bibitem{Geyi2011}
W.~Geyi, \emph{Foundations of Applied Electrodynamics}.\hskip 1em plus 0.5em
  minus 0.4em\relax John Wiley \& Sons, 2011.

\bibitem{TEAT-7231}
M.~Cismasu, D.~Tayli, and M.~Gustafsson, ``Stored energy based {3D} antenna
  analysis and design,'' Lund University, Department of Electrical and
  Information Technology, P.O. Box 118, S-221 00 Lund, Sweden, Tech. Rep.
  LUTEDX/(TEAT-7231)/1-18/(2014), 2014.

\bibitem{Geyi2014}
W.~Geyi, ``On stored energies and radiation {Q},'' \emph{arXiv preprint
  arXiv:1403.3129}, 2014.

\bibitem{Gustafsson+Nordebo2013}
M.~Gustafsson and S.~Nordebo, ``Optimal antenna currents for {Q},
  superdirectivity, and radiation patterns using convex optimization,''
  \emph{IEEE Trans. Antennas Propagat.}, vol.~61, no.~3, pp. 1109--1118, 2013.

\bibitem{Cismasu+Gustafsson2014a}
M.~Cismasu and M.~Gustafsson, ``Antenna bandwidth optimization with single
  frequency simulation,'' \emph{IEEE Trans. Antennas Propagat.}, vol.~62,
  no.~3, pp. 1304--1311, 2014.

\bibitem{Chu1948}
L.~J. Chu, ``Physical limitations of omnidirectional antennas,'' \emph{J. Appl.
  Phys.}, vol.~19, pp. 1163--1175, 1948.

\bibitem{Thal2006}
H.~L. Thal, ``New radiation {Q} limits for spherical wire antennas,''
  \emph{IEEE Trans. Antennas Propagat.}, vol.~54, no.~10, pp. 2757--2763, Oct.
  2006.

\bibitem{Thal2012}
------, ``{Q Bounds for Arbitrary Small Antennas: A Circuit Approach},''
  \emph{IEEE Trans. Antennas Propagat.}, vol.~60, no.~7, pp. 3120--3128, 2012.

\bibitem{Guillemin1963}
E.~A. Guillemin, \emph{Theory of linear physical systems}.\hskip 1em plus 0.5em
  minus 0.4em\relax New York: John Wiley \& Sons, 1963.

\bibitem{Harrington1968}
R.~F. Harrington, \emph{Field Computation by Moment Methods}.\hskip 1em plus
  0.5em minus 0.4em\relax New York: Macmillan, 1968.

\bibitem{Geyi+Jarmuszewski+Qi2000}
W.~Geyi, P.~Jarmuszewski, and Y.~Qi, ``The {F}oster reactance theorem for
  antennas and radiation {Q},'' \emph{IEEE Trans. Antennas Propagat.}, vol.~48,
  no.~3, pp. 401--408, Mar. 2000.

\bibitem{Cismasu+Gustafsson2014b}
M.~Cismasu and M.~Gustafsson, ``Multiband antenna {Q} optimization using stored
  energy expressions,'' \emph{IEEE Antennas and Wireless Propagation Letters},
  vol.~13, no. 2014, pp. 646--649, 2014.

\bibitem{Gustafsson+etal2014c}
M.~Gustafsson, J.~Frid\'en, and D.~Colombi, ``Antenna current optimization for
  lossy media with near field constraints,'' \emph{Antennas and Wireless
  Propagation Letters, IEEE}, 2014.

\bibitem{Brune1931}
O.~Brune, ``Synthesis of a finite two-terminal network whose driving-point
  impedance is a prescribed function of frequency,'' \emph{MIT J. Math. Phys.},
  vol.~10, pp. 191--236, 1931.

\bibitem{Alu+etal2007}
A.~Al{\`u}, M.~G. Silveirinha, A.~Salandrino, and N.~Engheta,
  ``Epsilon-near-zero metamaterials and electromagnetic sources: Tailoring the
  radiation phase pattern,'' \emph{Physical Review B}, vol.~75, no.~15, p.
  155410, 2007.

\bibitem{Gustafsson+Sjoberg2010a}
M.~Gustafsson and D.~Sj\"{o}berg, ``Sum rules and physical bounds on passive
  metamaterials,'' \emph{New Journal of Physics}, vol.~12, p. 043046, 2010.

\end{thebibliography}

\end{document}